\documentclass{emulateapj}
\usepackage{epsfig,graphicx}

\newcommand{\mc}{\multicolumn}
\newcommand{\expnt}[2]{\ensuremath{#1 \times 10^{#2}}}   
\newcommand{\gsim}{\gtrsim}

\newcommand{\chandra}{\textit{Chandra}}
\def\micron {\ensuremath{\mbox{ }\mu\mbox{{m}}}}
\newcommand{\um}{\micron}

\newcommand{\kms}{\ensuremath{\mbox{km}\,\mbox{s}^{-1}}}
\newcommand{\masyr}{\ensuremath{\mbox{mas}\,\mbox{yr}^{-1}}}

\newcommand{\sgr}{SGR~1900+14}
\newcommand{\axp}{1E~2259+586}

\newcommand{\snr}{SNR~G42.8+0.6}

\submitted{Accepted for publication in the Astronomical Journal}
\shortauthors{Kaplan et al.}
\shorttitle{Constraining the Proper Motions of Two Magnetars}

\begin{document}

\title{Constraining the Proper Motions of Two Magnetars}

\author{D.~L.~Kaplan\altaffilmark{1,2}, S. Chatterjee\altaffilmark{3}, B.~M.~Gaensler\altaffilmark{3},
  P.~O.~Slane\altaffilmark{4}, and C.~Hales\altaffilmark{3}}

\altaffiltext{1}{Pappalardo Fellow and Hubble Fellow; Kavli Institute for Astrophysics and Space
  Research and Department of Physics, Massachusetts Institute of
  Technology, Cambridge, MA 02139} 
\altaffiltext{2}{Current address: KITP, Kohn Hall, UCSB, Santa
  Barbara, CA 93106-4030; dkaplan@kitp.ucsb.edu.}
\altaffiltext{3}{Institute of Astronomy, School of Physics, The University of Sydney, NSW
  2006, Australia; schatterjee, bgaensler@usyd.edu.au, chales@physics.usyd.edu.au}
\altaffiltext{4}{Harvard-Smithsonian Center for Astrophysics, Cambridge, MA 02138;
  {slane@cfa.harvard.edu}.}

\begin{abstract}
We attempt to measure the proper motions of two magnetars --- the
soft gamma-ray repeater \sgr\ and the anomalous X-ray pulsar \axp\ ---
using two epochs of \chandra\ observations separated by $\sim 5\,$yr.  We perform extensive
tests using these data, archival data, and simulations to verify the
accuracy of our measurements and understand their limitations.   We
find 90\% upper limits on the proper motions of $54\,\masyr$ (\sgr)
and $65\,\masyr$ (\axp), with the limits largely determined by the
accuracy with which we could register the two epochs of data and by
the inherent uncertainties on two-point proper motions. 
We translate the proper motions limits into limits on the transverse
velocity using distances, and find $v_{\perp}<1300\,\kms$ (\sgr, for a
distance of 5$\,$kpc) and $v_{\perp}<930\,\kms$ (\axp, for a distance
of 3$\,$kpc) at 90\% confidence; the range of possible distances for
these objects makes a wide range of velocities possible, but it seems
that the magnetars do not have uniformly high space velocities of
$>3000\,\kms$.  Unfortunately, our proper motions also cannot
significantly constrain the previously proposed origins of these
objects in nearby supernova remnants or star clusters, limited as much
by our ignorance of ages as by our proper motions.

\end{abstract}


\section{Introduction}
The sources known as magnetars (see \citealt{wt06} for a review),
comprised of the observational classes soft gamma-ray repeaters (SGRs)
and anomalous X-ray pulsars (AXPs), are thought to be young neutron
stars with magnetic fields above $10^{14}$~G \citep{dt92,p92}.  Why
some neutron stars are magnetars (having magnetic field decay as their
primary energy source and usually emitting no detectable radio
pulsations) while others are ordinary rotation-powered radio pulsars
(with comparatively little X-ray emission; \citealt{km05} and
references therein) is a fundamentally unresolved issue, especially
since some radio pulsars have been discovered with field strengths
comparable to those of magnetars \citep[e.g.,][also see
\citealt{ggg+08}]{km05}. The proposed explanations run the gamut from
evolutionary sequences \citep[e.g.,][]{lz04,kaa06} and quiescent
states \citep[e.g.,][]{wt06,ghbb04} to differences in progenitor mass
\citep{gmo+05} or magnetic field orientation and geometry
\citep{zh00,kkm+03}.  \citet{dt92} have proposed that an initial short
birth period is responsible for the generation of the high magnetar
fields through an efficient large scale dynamo, and have also pointed
out that high magnetic fields can produce asymmetric neutrino emission
at birth, resulting in extreme space velocities ($>10^3\,{\rm
km}\,{\rm s}^{-1}$; also see \citealt{td93,td95,lai01b}).  Such large
velocities would also help explain the apparent offsets between some
SGRs and their putative natal supernova remnants
\citep[e.g.,][]{rkl94}.  These results for magnetars, taken with the
extensive investigation of rotation-powered pulsars
\citep[e.g.,][]{nr07}, imply that the initial spin periods, surface
magnetic fields, and birth kick velocities of NS all originate in
supernova core collapse processes, and the physics of these phenomena
are tightly interwoven.  {Magnetars, by virtue of their extreme
magnetic fields, may provide a direct probe of this interdependence.}
For example, if birth kicks are driven by asymmetric neutrino emission
mediated by high B-fields (a mechanism that may not be able to produce
the highest kick velocities; \citealt{lai01b}), then magnetars should have a
population velocity much higher than that of the radio pulsar
population.  However, a more critical investigation of some of the
claimed SGR/supernova remnant associations led \citet{gsgv01} to doubt
many associations and to conclude that magnetars as a class had
velocities $<500\,{\rm km}\,{\rm s}^{-1}$, comparable to the radio
pulsar population.

Such speculations were largely without concrete examination, as the
velocities of magnetars are difficult to measure.  As a class, they lack radio counterparts and so
cannot be used for traditional Very Long Baseline Interferometry, and
their spin-down properties are too noisy for measurement of ``timing''
proper motions.  This has changed recently, though.  The first
measurement of the direct proper motion (and hence space velocity,
with an assumed distance) of a magnetar was the work of
\citet{hcb+07}, who found a proper motion of $13.5\pm1.0\,{\rm
mas}\,{\rm yr}^{-1}$ ($\approx 212\,{\rm km}\,{\rm s}^{-1}$ at a
distance of $3.5\,$kpc) for \object[XTE J1810-197]{XTE~J1810$-$197}.  This was  made
possible by the detection of a bright, transient radio
counterpart\footnote{\citet{hcb+07} also used infrared measurements of
  XTE~J1810$-$197 to constrain the proper motion, and found a much
  coarser but consistent result.}
\citep{hgb+05}.

In this paper, we attempt to further address this issue by trying to
measure the proper motions of two magnetars, \object[SGR 1900+14]{\sgr}\ and \object[PSR J2301+5852]{\axp}.  In the
absence of steady or widespread radio counterparts, we try to measure proper motions
in the X-rays: while \axp\ has an infrared counterpart \citep{htvk+01}
and would also be amenable to a measurement using data at those
wavelengths, we try to use a technique applicable to objects without
infrared counterparts, although we anticipate  infrared data being used to
check our measurement \citep*{cbk08}.

Not only can our data provide valuable constraints on the velocity
distribution of magnetars, but they can also potentially address
associations between the magnetars and other objects (supernova
remnants and massive star clusters).  Such associations are valuable
for the constraints they can place on the ages, distances, and
progenitor masses of magnetars \citep[e.g.,][]{vhl+00,gsgv01,mcc+06}
We discuss this in more detail
below.

The structure of this paper is as follows.  First, in
\S~\ref{sec:sources} we discuss the two targets of this analysis,
\sgr\ and \axp.  We then discuss the archival data and our new
\textit{Chandra X-ray Observatory} Advanced CCD Imaging Spectrometer
(ACIS; \citealt{gbf+03}) observations in \S~\ref{sec:obs}.
In \S~\ref{sec:xray}, we present a detailed analysis of those data,
along with archival data used to test our techniques and simulations.
We present our proper motion measurements in \S~\ref{sec:pm}, along
with comparison to other similar measurements and the prospects for
improvement.  Finally, we give our discussion and conclusions in
\S~\ref{sec:disc}.

\subsection{\sgr\ and \axp}
\label{sec:sources}
The third SGR discovered \citep*{mgg79}, \sgr\ was identified as a
persistent X-ray source by \citet{hlv+96}.  Subsequent observations
identified  the 5.2-s pulse period \citep{hlk+99} with
significant spin-down \citep{ksh+99} implying a large dipolar magnetic
field.  The magnetar nature of this source was confirmed
by the giant flare detected on 1998~August~27
\citep{hcm+99,td95,td96}.

\citet{vkfg94} and \citet{kfm+94} suggested that \sgr\ could be
associated with the nearby supernova remnant (SNR) G42.8+0.6, based
largely on position coincidence and the possible association of a
number of other AXPs and SGRs with SNRs.  However, \citet{lx00} called
the association into question when they discovered a relatively young
($10^4$--$10^5$~yr old) radio pulsar near both the SGR and the SNR,
and \citet{gsgv01} calculated a 4\% chance of random alignment between
the SGR and SNR.  This chance is even greater if one accepts that
supernova explosions would naturally tend to be clustered, so \sgr\
appearing near \snr\ (and other SNRs; \citealt{kkfvk02}) is not
unlikely.  If the association between \sgr\ and \snr\ were real, \sgr\
would need a significant transverse velocity to make it from the
explosion site to its current location outside the SNR's radio shell
\citep{hkw+99b}.  \sgr\ is $\approx 15\arcmin$ from the center of the
SNR (at approximately $\alpha_{\rm J2000}=19^{\rm h}07^{\rm m}01^{\rm
s}$, $\delta_{\rm J2000}=+09\degr04\arcmin10\arcsec$ from
low-frequency radio images, with an uncertainty of $\pm30\arcsec$;
\citealt{kkfvk02}).  This implies a proper motion of
$90\tau_{10}^{-1}\,\masyr$ or a transverse velocity
$v_{\perp}\approx 2100d_5 \tau_{10}^{-1}\,\kms$, where the distance to
the system is $5d_5$~kpc, and the age\footnote{Both the age and
distance of the \sgr\ are highly uncertain.  The spin-down age
$P/2\dot P$ is about 1$\,$kyr, but the spin-down rates of magnetars
can easily vary by factors of several \citep{wt06}, thus making the
spin-down age an unreliable estimator of the true age.  The distance
to \sgr\ itself is largely unconstrained.} is $10\tau_{10}$~kyr.  This is a
large velocity for a neutron star \citep[e.g.,][]{hllk05}, but not
unheard of \citep{cvb+05}.

\begin{deluxetable*}{l c l l c c c c}
\tablewidth{0pt}
\tablecaption{Summary of \chandra\ Observations \label{tab:obs}}
\tablehead{
\colhead{Target} & \colhead{Epoch} & \colhead{ObsID} & \colhead{Date} & \colhead{MJD} & \colhead{Aimpoint}
& \colhead{Exposure} & \colhead{Rotation}\\
& & & & & & \colhead{(ksec)} & \colhead{(deg)}
}
\startdata
\sgr & 1 & \dataset[ADS/Sa.CXO#obs/1954]{1954} &2001-Jun-17 & 52077 & ACIS-I & 29.8 & 148.6\\
     & 2 & \dataset[ADS/Sa.CXO#obs/6731]{6731} &2006-Jun-04 & 53891 & ACIS-I & 24.6 & 134.0\\
\axp & 1 &\dataset[ADS/Sa.CXO#obs/725]{\phn725}&2000-Jan-12 & 51555 & ACIS-S & 18.9 & 312.8\\
     & 2 & \dataset[ADS/Sa.CXO#obs/6730]{6730}&2006-May-09 & 53864 & ACIS-S & 24.8 & \phn73.7\\
\enddata
\tablecomments{The rotation given is the nominal roll angle of the
  spacecraft, defined by the \texttt{ROLL\_NOM} header keyword.}
\end{deluxetable*}

An alternative association has been proposed for \sgr.  \citet{vhl+00}
identified a cluster of massive stars, with several M supergiant
members and $>10$ other stars, about $12\arcsec$ away from the \sgr.
This has been viewed as a much more secure association for \sgr, since
such clusters are rarer and it is much closer on the sky, and
\citet{wrrd+08} claim that the infrared ring found around \sgr\ could
only be powered by the cluster stars.  The possible inconsistency
between the extinction to \sgr\ and the cluster \citep{kkfvk02} may
not be significant, as demonstrated by \citet{wrrd+08}.  This cluster
is at 12--15$\,$kpc, so if the association is real then \sgr\ is at a
distance of $d_5\approx 3$.

The anomalous X-ray pulsar \axp\ was the first member of its class to
be discovered, when \citet{fg81} identified it as an X-ray point
source in the center of the SNR \object[CTB 109]{CTB~109} that had coherent $7$-s
pulsations.  The association with the SNR seems secure \citep{gsgv01},
as the positional coincidence is very good ($<3\arcmin$) compared to
the local density of SNRs.  Based on interactions between the SNR and
\ion{H}{2} regions with measured distances, \citet*{kuy02} find a
distance\footnote{There is some uncertainty about the association and
hence the distance to \axp.  \citet{dvk06} find a distance to \axp\ of
$7.5\pm1.0\,$kpc using the ``red clump'' method.  They argue that
CTB~109 could also plausibly be at this distance too, based on a
reinterpretation of radial velocity data, but there are uncertainties
associated with this analysis.} of $\approx
3\,$kpc.

\section{Observations}
\label{sec:obs}
We examined the \chandra\ archive for suitable observations of
magnetars.  We found data on \sgr\ and \axp\ that we used for the
first epochs of our proper motions studies. These data were taken with
the ACIS instrument in the full-frame mode and with no grating, in
contrast to many other observations of magnetars that use subarrays,
continuous clocking, or gratings to get better timing or spectral
resolution.

We re-observed \sgr\ and \axp\ with \chandra, with the new data
serving as second epoch observations to try to measure or constrain
their proper motions.  For the second epochs, we chose to generally
replicate the first epochs in detector choice, aimpoint, and exposure
time.  One of the priorities was to detect as many other sources as
possible besides the \sgr/\axp: as discussed below, our accuracy is
limited by the number and quality of the reference sources.
In Table~\ref{tab:obs} we give
the details of both the archival observations and our new ones.  Note
that the rotations of the new and old observations are in general
different.  For \sgr, which used the square ACIS-I detector, this made
no difference, but for \axp\ the reference sources changed somewhat
between epochs, giving us a smaller set than we would have liked: see
discussion in \S~\ref{sec:axp}.

\section{X-ray Astrometry}
\label{sec:xray}
Our general analysis technique is to (1) measure the positions of a
number of field X-ray sources (``reference sources'') common to
both epochs; (2) measure the position of the object of interest (\sgr\
or \axp) in both epochs; (3) use the reference sources to determine a
transformation between the two observations; and (4) apply that
transformation to the object of interest, finding the position
difference (if any) between the two epochs that we can then convert to
a proper motion.  This generally echoes the techniques of
\citet{hb06}, \citet{wp07}, and \citet{mphs07,mph+07}, although like
\citet{mphs07,mph+07} we prefer to use many reference sources and not
just a few.

We must determine the best way to carry out each step, and assess the
uncertainties associated with each.  For this reason, we discuss each
step at length below, in many cases using data from other observations
and/or simulations to assess accuracy.

In the discussion below, we work as much as possible in the sky pixel
$(x_{\rm sky},y_{\rm sky})$ frame.  This is a frame constructed by the
\chandra\ event processing that rotates the detector $(x_{\rm
det},y_{\rm det})$ frame to have $y_{\rm sky}$ pointing north and
$x_{\rm sky}$ pointing west.  While in general we prefer to stay in a
frame with as little processing as possible, which would argue for
retaining the detector frame, the continuous dither of \chandra\ (in a
Lissajous pattern with amplitude $\pm8\arcsec$ and periods of 707$\,$s
and 1000$\,$s along the two axes) makes that difficult.  The
conversion from detector to sky frame involves correcting for that
dither, for the nominal orientation of the spacecraft, and for any
other position excursions, but this aspect reconstruction is of
sufficient quality that it does not limit our accuracy.

\subsection{Reference Sources}
\label{sec:ref}
For our main position-finding algorithm, we chose the
CIAO task \texttt{wavdetect} \citep{fkrl02}, which uses a wavelet-based
detection algorithm.  This echoes the choices made by such large X-ray
surveys as the \chandra\ Multiwavelength Plane (ChaMPlane) Survey
\citep{hvdbs+05} and the \chandra\ Multiwavelength Project (ChaMP;
\citealt{kkw+07}).  In those cases the algorithm choice was
made for optimum source detection, with astrometry only as a
secondary goal, but they did find that \texttt{wavdetect} produced the
most reliable astrometry.  We verified this choice by comparing the
results of \texttt{wavdetect}, the sliding-cell task
\texttt{celldetect}, and our own implementation of an interactive centroiding task
that operates on the raw event lists, and found that
\texttt{wavdetect} gave the smallest scatter in the various tests
described below.  

As we were not overly concerned with detecting every possible source
or measuring such source properties as count-rate, we ran
\texttt{wavdetect} on the raw level 2 event files returned from the
\chandra\ archive without any additional processing (e.g., exposure
map creation), using wavelet scales of $1,2,4,8$~pixels.  We did test
whether removing sub-pixel randomization\footnote{See
\url{http://cxc.harvard.edu/ciao/threads/acispixrand/}.} made a
difference in the positions, and we find no systematic difference.

However, the uncertainties reported by \texttt{wavdetect} are
systematically low, especially for off-axis sources.  We therefore
investigated the uncertainty models used by the ChaMP and ChaMPlane
projects.  Independent of the analyses of \citet{hvdbs+05} and
\citet{kkw+07}, which relied on \texttt{MARX} (Model of AXAF Response
to X-rays) and \texttt{SAOsac} (a raytracing code developed for
\chandra)
simulations of sources, we wished to see how reliable the position
measurement and uncertainty estimation of \texttt{wavdetect} are.
Therefore we took a number of relatively long ACIS-I exposures from
the archive with a large number of point sources.  We then split those
exposures into two halves by time, effectively making two
sub-exposures that are otherwise identical, and processed each with
\texttt{wavdetect} in the same manner.  This way we can examine how
the positions from the first half agree with those from the second
half: since the two sub-exposures are from the same observation, the
positions should agree.  Of course the sources in each sub-exposure
with have fewer counts than the sources in the total exposure, but
there are still a range of count levels and off-axis angles.

We first did this comparing the results of \texttt{wavdetect} with
\texttt{celldetect} and an interative centroiding algorithm that we implemented,
and found \texttt{wavdetect} to have the lowest dispersion.  We then
varied the energy range of our source extraction and found that
including event energies from 0.3~keV to 7~keV optimized the
signal-to-noise of the extraction.

We then compared the position differences between the first and second
sub-exposures with the position uncertainties returned by
\texttt{wavdetect}.  We found that \texttt{wavdetect} significantly
underestimated the uncertainties, especially at off-axis angles
$\theta>3.5\arcmin$.  The comparison gave a reduced $\chi^2$ of 4.31
for 1450~degrees-of-freedom (DOF), and the shape of the distribution
is clearly broader than the expected distribution, with more points at
$>2\sigma$.  In contrast, both the ChaMP and ChaMPlane models work
well, giving reduced $\chi^2$ values of $1.33$ and 1.02.  The
uncertainty from the ChaMPlane model exceeds that from the  ChaMP model slightly at large and small
off-axis angles, leading to the slightly lower $\chi^2$ for ChaMPlane.

Overall, both the ChaMP and ChaMPlane uncertainty models appear to work.  Both are based on
simulations using \texttt{MARX}, and so
the agreement is not surprising, but since they agree with each other
and also agree with the actual data presented here (see
\S~\ref{sec:tform}) we feel comfortable using either.  The functional
form of the ChaMP model is somewhat simpler and conforms more to the
expected relation between uncertainty $\sigma_{x,y}$ and
signal-to-noise ratio, with $\sigma_{x,y} \propto C^{-0.46}$ ($C$ is
the number of counts above the background) compared to the expected $\sigma_{x,y}
\propto C^{-0.5}$ (i.e.\ $\sigma_{x,y} \propto (S/N)^{-1/2}$).  In
contrast, the ChaMPlane model is a more complicated function of $C$
involving powers of the logarithm of $C$.  Therefore we will use the
ChaMP model (see eq.~14 of \citealt{kkw+07}):
\begin{equation}
\sigma_{x,y}=0.77\, C^{-0.46}\, 10^{0.11 \theta}\,\mbox{pixels}
\label{eqn:dposs}
\end{equation}
with $\sigma_{x,y}$ as the $x$ or $y$ uncertainty (converted from the
radial uncertainty in \citealt{kkw+07} by dividing by 1.515; see
\citealt*{lmb76}), and $\theta$ in arcminutes.  Note that this is only
valid for low count levels ($C<131$) as is the case here but not for
\sgr\ or \axp.

\subsection{\sgr}
\label{sec:magpos}
For \sgr, we are in a very different regime
than for the reference sources.  Instead of relatively faint sources
at a variety of off-axis angles, we have a bright source at
$\theta<0.5\arcmin$.  A different approach to position measurement and
uncertainties than that described in \S~\ref{sec:ref} is needed.

\begin{deluxetable*}{l c c c c c c c c c}
\tabletypesize{\footnotesize}
\tablewidth{0pt}
\tablecaption{Source Positions for \sgr\label{tab:ref}}
\tablehead{
\colhead{ID} & \mc{4}{c}{Epoch 1} && \mc{4}{c}{Epoch 2} \\ \cline{2-5} \cline{7-10}
 & \colhead{$x_{\rm sky}$}  & \colhead{$y_{\rm sky}$} & \colhead{Counts}
& \colhead{$\theta$ (arcmin)} & & \colhead{$x_{\rm sky}$}  & \colhead{$y_{\rm sky}$}& \colhead{Counts}
& \colhead{$\theta$ (arcmin)}
}
\startdata
\phn X & $4075.5 \pm 0.0$& $4067.4 \pm 0.0$ &  5232.3 & 0.3 && $4109.7 \pm 0.0$& $4022.9 \pm 0.0$ &  3805.6 & 0.3 \\
\phn 1 & $3991.4 \pm 0.5$& $4425.6 \pm 0.5$ & \phn\phn 15.5 & 2.8 && $4025.8 \pm 0.5$& $4380.5 \pm 0.5$ & \phn\phn 11.7 & 2.8 \\
\phn 2 & $4080.5 \pm 0.4$& $4494.6 \pm 0.4$ & \phn\phn 30.5 & 3.3 && $4114.7 \pm 0.6$& $4448.8 \pm 0.6$ & \phn\phn 12.7 & 3.2 \\
\phn 3 & $4159.6 \pm 1.0$& $4707.2 \pm 1.0$ & \phn\phn 10.5 & 5.0 && $4192.8 \pm 1.1$& $4662.8 \pm 1.1$ & \phn\phn\phn 8.1 & 5.0 \\
\phn 4 & $4385.9 \pm 0.4$& $4822.8 \pm 0.4$ & \phn 138.2 & 6.4 && $4419.6 \pm 0.5$& $4779.3 \pm 0.5$ & \phn\phn 95.3 & 6.4 \\
\phn 5 & $4409.6 \pm 0.7$& $4421.9 \pm 0.7$ & \phn\phn\phn 9.6 & 3.7 && $4442.6 \pm 1.3$& $4378.2 \pm 1.3$ & \phn\phn\phn 2.9 & 3.7 \\
\phn 6 & $4025.4 \pm 1.5$& $4785.1 \pm 1.5$ & \phn\phn\phn 6.3 & 5.7 && $4059.5 \pm 1.0$& $4743.5 \pm 1.0$ & \phn\phn 15.8 & 5.7 \\
\phn 7 & $4383.0 \pm 1.7$& $4655.8 \pm 1.7$ & \phn\phn\phn 3.5 & 5.2 && $4418.8 \pm 1.4$& $4612.8 \pm 1.4$ & \phn\phn\phn 5.7 & 5.2 \\
\phn 8 & $4395.0 \pm 1.9$& $5093.7 \pm 1.9$ & \phn\phn 18.6 & 8.5 && $4425.4 \pm 2.3$& $5053.4 \pm 2.3$ & \phn\phn 12.5 & 8.6 \\
\phn 9 & $3638.7 \pm 0.6$& $4492.5 \pm 0.6$ & \phn\phn 29.0 & 5.0 && $3674.1 \pm 0.8$& $4446.7 \pm 0.8$ & \phn\phn 15.8 & 4.9 \\
10 & $3997.8 \pm 0.5$& $4280.2 \pm 0.5$ & \phn\phn\phn 5.7 & 1.7 && $4032.3 \pm 0.5$& $4235.4 \pm 0.5$ & \phn\phn\phn 6.8 & 1.7 \\
11 & $3055.8 \pm 1.9$& $4475.3 \pm 1.9$ & \phn\phn 23.9 & 9.1 && $3090.0 \pm 1.9$& $4426.7 \pm 1.9$ & \phn\phn 24.1 & 9.0 \\
12 & $3112.3 \pm 1.0$& $4680.9 \pm 1.0$ & \phn 128.4 & 9.4 && $3148.0 \pm 2.0$& $4635.6 \pm 2.0$ & \phn\phn 24.8 & 9.3 \\
13 & $3319.6 \pm 1.3$& $4314.9 \pm 1.3$ & \phn\phn 13.8 & 6.6 && $3355.5 \pm 1.2$& $4270.0 \pm 1.2$ & \phn\phn 16.7 & 6.5 \\
14 & $4313.9 \pm 0.4$& $4043.5 \pm 0.4$ & \phn\phn\phn 9.7 & 1.8 && $4348.5 \pm 0.5$& $3999.2 \pm 0.5$ & \phn\phn\phn 7.7 & 1.9 \\
15 & $4398.5 \pm 0.6$& $4043.0 \pm 0.6$ & \phn\phn\phn 7.7 & 2.5 && $4433.8 \pm 0.7$& $3998.2 \pm 0.7$ & \phn\phn\phn 5.8 & 2.6 \\
16 & $4706.4 \pm 0.9$& $3894.6 \pm 0.9$ & \phn\phn 14.0 & 5.3 && $4739.4 \pm 1.1$& $3848.5 \pm 1.1$ & \phn\phn\phn 8.9 & 5.3 \\
17 & $4781.0 \pm 0.7$& $3963.5 \pm 0.7$ & \phn\phn 35.0 & 5.7 && $4814.3 \pm 1.0$& $3918.2 \pm 1.0$ & \phn\phn 15.3 & 5.8 \\
18 & $4694.9 \pm 1.1$& $4248.4 \pm 1.1$ & \phn\phn\phn 9.1 & 5.1 && $4730.5 \pm 1.0$& $4200.2 \pm 1.0$ & \phn\phn 10.1 & 5.1 \\
19 & $3839.5 \pm 0.6$& $3917.6 \pm 0.6$ & \phn\phn\phn 7.7 & 2.6 && $3873.1 \pm 0.5$& $3872.6 \pm 0.5$ & \phn\phn 10.9 & 2.5 \\
20 & $3931.1 \pm 0.5$& $3957.6 \pm 0.5$ & \phn\phn\phn 8.7 & 1.8 && $3965.4 \pm 0.4$& $3912.9 \pm 0.4$ & \phn\phn\phn 9.8 & 1.7 \\
21 & $4005.7 \pm 0.4$& $3945.3 \pm 0.4$ & \phn\phn\phn 8.6 & 1.4 && $4040.1 \pm 0.4$& $3900.6 \pm 0.4$ & \phn\phn\phn 8.7 & 1.4 \\
22 & $4046.3 \pm 0.3$& $3940.9 \pm 0.3$ & \phn\phn 13.6 & 1.3 && $4080.3 \pm 0.5$& $3896.7 \pm 0.5$ & \phn\phn\phn 5.8 & 1.3 \\
23 & $3582.5 \pm 0.9$& $3748.2 \pm 0.9$ & \phn\phn 12.9 & 5.1 && $3617.3 \pm 1.0$& $3703.3 \pm 1.0$ & \phn\phn 10.4 & 5.0 \\
24 & $3925.2 \pm 1.1$& $3545.0 \pm 1.1$ & \phn\phn\phn 7.4 & 4.7 && $3958.9 \pm 0.8$& $3499.8 \pm 0.8$ & \phn\phn 13.9 & 4.7 \\
25 & $4125.5 \pm 1.0$& $3069.3 \pm 1.0$ & \phn\phn 72.4 & 8.4 && $4158.6 \pm 1.1$& $3024.0 \pm 1.1$ & \phn\phn 53.8 & 8.5 \\
26 & $4207.8 \pm 1.4$& $3368.5 \pm 1.4$ & \phn\phn\phn 8.5 & 6.0 && $4241.9 \pm 1.0$& $3323.4 \pm 1.0$ & \phn\phn 16.3 & 6.1 \\
27 & $4255.9 \pm 0.9$& $3715.3 \pm 0.9$ & \phn\phn\phn 4.8 & 3.4 && $4289.4 \pm 0.7$& $3670.6 \pm 0.7$ & \phn\phn\phn 7.8 & 3.4 \\

\enddata
\tablecomments{Positions are in sky pixel coordinates, $\theta$ is the
  off-axis angle, and source X is \sgr.}
\end{deluxetable*}

The ACIS-I images of \sgr\ have the source at an off-axis angle of
$20\arcsec$, and with about 5000~counts.  Technically, then, the ChaMP
uncertainty models do not apply (they are only valid for
$<2000$~counts).  Even so, the ChaMP model for high numbers of counts
does not agree with what we would naively expect, with $\sigma_{x,y}
\propto C^{-0.2}$.  We therefore did our own \texttt{SAOsac/MARX}
(using the \chandra\ Ray Tracer ChaRT as an interface to the SAOsac
raytracing code; \citealt{ckj+03}) simulations of this source.

We simulated a bright point-source at the correct off-axis location
with the approximate spectrum of \sgr: a power-law with photon index
$\Gamma=1.9$ and absorption with $N_{\rm H}=\expnt{2}{22}\,\mbox{cm}^{-2}$ \citep{met+06}.  We simulated a source with
700,000$\,$counts, many more
counts than are actually detected in our exposures of \sgr, so that we could
divide this exposure into sub-exposures as above for comparison.  We
made a series of sub-exposures, dividing the 700,000 total counts into
12 sub-exposures with 56,000~counts, 25 sub-exposures with
27,000~counts, and so on, down to 300 sub-exposures with 2200~counts.
We then ran \texttt{wavdetect} on each of the sub-exposures in a
series and compared all of the positions.  In contrast to the
situation with  the reference
sources, where the raw \texttt{wavdetect} uncertainties underpredicted
the uncertainties, here they did well.  In all cases the comparisons
yielded reduced $\chi^2$ values consistent with 1.0.  Fitting the
uncertainty as a function of number of counts, we find a relation
close to the expected one, with $\sigma_{x,y}\propto C^{-0.47}$.  

To confirm this, we also divided up the real exposures of \sgr\ into
10 sub-exposures of $\approx 500$~counts each.  Each position as
measured by \texttt{wavdetect} had an uncertainty of $0.04$~pixels,
and again the positions of the sub-exposures were all consistent with
each other within the uncertainties.  

\begin{figure}
\plotone{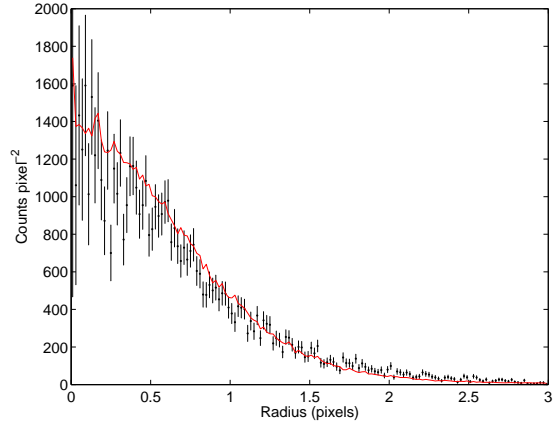}
\caption{Radial profile of \sgr\ (points) compared to that of a
  \texttt{MARX} simulation with 30 times the number of counts (solid
  line) that does not include the effects of pileup.  The agreement
  between the data and the model illustrates the small effect of
  pileup on the radial profile of \sgr.
}
\label{fig:rpsgr}
\end{figure}

A final concern is pileup.  With a count-rate of $\sim 0.17\,\mbox{s}^{-1}$, the expected pileup fraction is $\sim 20$\%, with
  $\sim0.5$~event  expected in each 3.2~s frame.  This means that
  our data are affected by pileup such that spectral estimation will
  not be robust, but at this pileup level the effect on astrometry is
  minimal.  The spatial
  distribution of events is still largely consistent with the
  point-spread function (Fig.~\ref{fig:rpsgr}).  Our \texttt{MARX}
  simulations here did not account for pileup, but even so they were
  consistent with our examination of the sub-exposures of the real
  data.  Overall, then, \texttt{wavdetect} positions and uncertainties
  seem sufficient for observations of \sgr.

\subsection{\axp}
For \axp, we could not use the results of \texttt{wavdetect}.  The
pileup for this source is severe enough that the X-ray image has a
central hole, where so many X-ray photons landed that all were judged
to be cosmic rays.  We also have issues at large radii: beyond a radius of $\approx 4\arcsec$, there is
extended emission from a combination of the SNR in which \axp\ is
embedded, as well as a halo from dust scattering \citep{pkw+01}.  

We initially tried doing simple centroids on the event data, where we
used events in an annulus whose inner radius varied from 0 to 5~pix
(to eliminate the central hole) and whose outer radius varied from
15~pix to 100~pix.  For data with no central hole, iterating this
process will converge on the correct center (this was one of the
methods we used in \S~\ref{sec:magpos}).  
However, we found that this method gave results
that did not converge: changing the outer radius led to
systematic changes in the centroid position.  This was partially due
to asymmetries in the extended X-ray emission, but more important were
the effects of non-uniform exposure over the ACIS CCDs. Together,
these effects led to variations (both smooth and discrete) of the
overall count level over the image, and when the annuli that we chose
included those portions, they skewed the resulting centroid.
 Simulating
the data in \texttt{MARX}, where we used similar exposure maps,
reproduced the wandering centroid, although not to a high enough
precision that the \texttt{MARX} data would allow us to correct the
observations.

\begin{figure}
\plotone{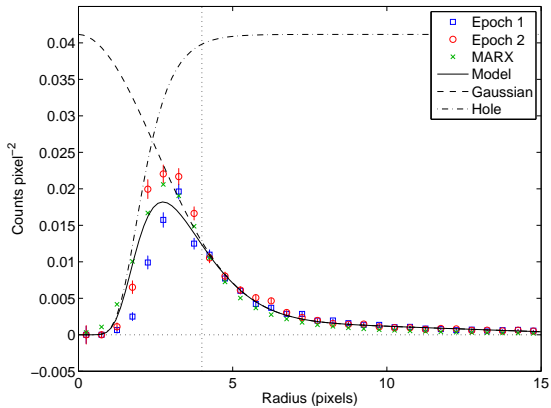}
\caption{Radial profiles of \axp\ (squares and circles for epochs 1 and
  2) compared to that of a \texttt{MARX} simulation that includes
  pileup (x's).  We fit the data outside of a radius of $4\,$pix
  (vertical dotted line) with a Gaussian (dashed line) multiplied by a
  hole (dot-dashed line; the hole has the form $\tanh(r/r_0)^{10}$ with
  $r_0=1.25\,$pix); the full model is the solid line.  }
\label{fig:rpaxp}
\end{figure}

\begin{deluxetable*}{l c c c c c c c c c}
\tabletypesize{\footnotesize}
\tablewidth{0pt}
\tablecaption{Source Positions for \axp\label{tab:refaxp}}
\tablehead{
\colhead{ID} & \mc{4}{c}{Epoch 1} && \mc{4}{c}{Epoch 2} \\ \cline{2-5} \cline{7-10}
 & \colhead{$x_{\rm sky}$}  & \colhead{$y_{\rm sky}$} & \colhead{Counts}
& \colhead{$\theta$ (arcmin)} & & \colhead{$x_{\rm sky}$}  & \colhead{$y_{\rm sky}$}& \colhead{Counts}
& \colhead{$\theta$ (arcmin)}
}
\startdata
Y & $4141.7\pm0.1$ & $4171.9\pm0.1$ &\nodata & 0.7 && $4117.0\pm0.1$ &
$4070.7\pm0.1$ & \nodata & 0.3\\
1 & $4093.8 \pm 0.4$& $3567.5 \pm 0.4$ & \phn 64.6 & 4.3 && $4069.9 \pm 0.3$& $3466.8 \pm 0.3$ &  157.7 & 5.2 \\
2 & $4536.1 \pm 0.2$& $3899.2 \pm 0.2$ &  295.7 & 4.0 && $4510.9 \pm 0.5$& $3798.9 \pm 0.5$ & \phn 26.6 & 4.2 \\
3 & $4159.5 \pm 0.3$& $3799.6 \pm 0.3$ & \phn 38.7 & 2.5 && $4135.0 \pm 0.4$& $3699.7 \pm 0.4$ & \phn 21.6 & 3.3 \\
4 & $4270.1 \pm 0.2$& $4755.3 \pm 0.2$ &  330.9 & 5.6 && $4245.3 \pm 0.3$& $4654.4 \pm 0.3$ &  103.4 & 4.7 \\
5 & $4350.6 \pm 0.4$& $4446.1 \pm 0.4$ & \phn 28.9 & 3.5 && $4326.3 \pm 0.3$& $4344.7 \pm 0.3$ & \phn 47.6 & 2.8 \\
6 & $4273.3 \pm 1.2$& $4961.5 \pm 1.2$ & \phn 24.4 & 7.2 && $4246.7 \pm 1.0$& $4857.9 \pm 1.0$ & \phn 20.8 & 6.4 \\

\enddata
\tablecomments{Positions are in sky pixel coordinates, $\theta$ is the
  off-axis angle, and source Y is \axp.}
\end{deluxetable*}

Instead, we followed \citet{htvk+01} and explicitly used a local,
symmetric function to find the position of \axp.  This is similar to
what we did with \sgr, where \texttt{wavdetect} found the position
that maximized the overlap between the data and a wavelet function,
but for \axp\ we use our own kernel function and our own routines to
find the position.  For the kernel we used a Gaussian function
multiplied by a hole, meant to mimic the shape of the piled-up image.
The function was radially symmetric.  \citet{htvk+01} used a
hyperbolic tangent for their ``hole'' function: we found that this did
not decline quite quickly enough to match our data, and instead used
$\tanh(r/r_0)^{10}$, where $r_0$ is the radius scale and the exponent
serves to increase the sharpness of the hole; the results are not
sensitive to the exponent as long as it is $\gsim 4$.  A fit to the
data from both epochs (averaged in azimuth), along with a
\texttt{MARX} simulation that includes pileup, is shown in
Figure~\ref{fig:rpaxp}.  To make that fit, we fit the Gaussian to the
data, the \texttt{MARX} model with pileup, and the \texttt{MARX} model
without pileup, all for radii of $\geq4\,$pix.  At those radii the
data and the two models all agree reasonably well.  Within $4\,$pix
the effects of pileup become more severe, and the \texttt{MARX} model
is not able to accurately predict the data, but the agreement is
qualitatively reasonable.  We found a hole radius $r_0\approx
1.25\,$pix, but again the fit was not extremely sensitive to this
value.  The goal of this model is to give a good qualitative
representation of the data, not to fit it in great detail.  

To actually find the position of \axp, instead of fitting the data,
which is sensitive to the details of the model as well, we  prefer to
cross-correlate a fixed model with the data.  Our fixed model is that
shown in Figure~\ref{fig:rpaxp}:
\begin{equation}
F(r)=e^{-r^2/2\sigma^2} \tanh^{10}\left(\frac{r}{r_0}\right)
\label{eqn:hole}
\end{equation}
where we found $\sigma=2.5\,$pixels and $r_0=1.25\,$pixels gave good results.  This
function is radially symmetric, and by cross-correlating rather than
fitting we can ignore issues of normalization and the variations in
the wings of the radial profile.  Above, we could
have just fit the data to a Gaussian function only for radii greater
than $4\,$pixels, but this would not have allowed the
cross-correlation.  Using a Gaussian kernel function with no hole
but just excluding a portion of the data performed poorly, since the
sharp ``horns'' of the data outside of the hole can cause local maxima
in the cross correlation if the kernel function has a single peak like
a Gaussian.

\begin{figure}
\plotone{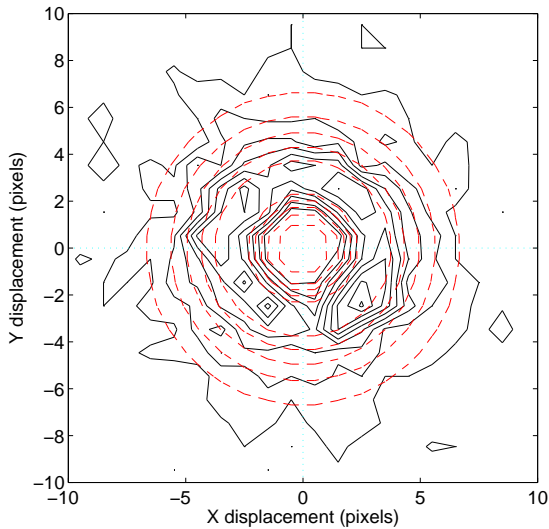}
\caption{Cross-correlation of the first epoch of data for \axp\ (black
  solid contours) with the model from Eqn.~\ref{eqn:hole} (red dashed
  contours).  The model and data have been shifted to the origin based
  on the result of the cross-correlation.  The results for epoch 2 are
  similar.}
\label{fig:xcorr}
\end{figure}

We cross-correlated  images of the filtered event files created
from both the data and the \texttt{MARX} simulations with the
model (Fig.~\ref{fig:xcorr}).  The cross-correlation results were smooth and well-behaved,
and we interpolated between our grid points to find the best-fit offset.
We made sure that the fitted position did not depend on the
details of the cross-correlation: we varied the binning, the outer
radius, and the parameters of the model ($\sigma$ and $r_0$), and in
all cases the resulting variations in the position were $<0.05\,$pix.
As a further check, we shifted the event positions by random amounts of up to
$1\,$pix, and found that we were able to recover the shifts to an
accuracy of $<0.1\,$pix; this test also verified that our result did
not depend on the origin of the binning.  Based on these tests, as
well as additional \texttt{MARX} simulations, we estimate that our
cross-correlation is accurate to $\pm0.1\,$pix in each coordinate.

\begin{figure}
\plotone{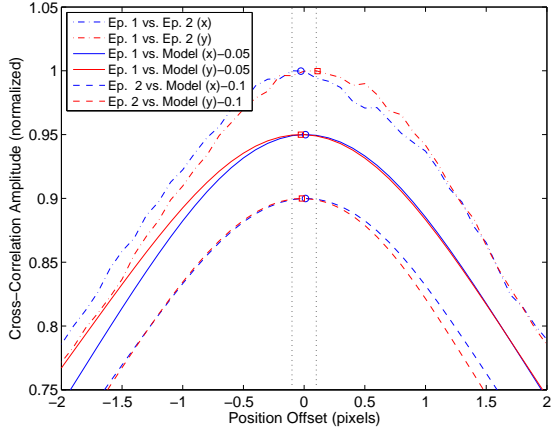}
\caption{Cross-correlation results for \axp.  We show one-dimensional
  cuts through the peak of the cross-correlation surface, which we use
  to identify the best-fit positions in $x$ (blue curves) and $y$ (red
  curves).  For each cross-correlation (epoch 1 vs.\ epoch 2:
  dot-dashed line; epoch 1 vs.\ the model: solid line; epoch 2 vs.\
  the model: dashed line; in all cases the model is from
  Eqn.~\ref{eqn:hole}) we interpolated to find the maximum value of
  the cross-correlation, which we identify as the best-fit position,
  and the cross-correlations have been offset vertically for clarity.
  The data plotted here have had our nominal positions subtracted,
  giving offsets near 0.  They are not quite at zero as the nominal
  positions were the averages of cross-correlations done with a range
  of random offsets added to mitigate the effects of binning, but the
  differences are consistent with our claimed uncertainty of
  0.1$\,$pix (vertical dotted lines).}
\label{fig:xcorroffsets}
\end{figure}

As another test, we also cross-correlated the binned images from
epochs 1 and 2 against each other, rather than against a smooth
kernel.  In general the results of this test were similar, where we
could recover an input shift at a level of $<0.1\,$pix, and the
offsets agreed to a similar precision.  We plot the results of all
cross-correlations (epoch 1 vs.\ model, epoch 2 vs.\ model, and epoch
1 vs.\ epoch 2) in Figure~\ref{fig:xcorroffsets}, where we plot the
results relative to the best-fit positions given in
Table~\ref{tab:refaxp}.  The cross-correlations that we plot do not
agree exactly with the positions in Table~\ref{tab:refaxp} as those
positions were derived using multiple cross-correlations with random
offsets used to eliminate the effects of binning, but the agreement is
better than the quoted $\pm0.1\,$pix uncertainty.

For a final check, we examined the position of the readout streak on
the two epochs.  This streak comes about because the ACIS detector has
no shutter, so the CCDs receive photons even as the charge is being
clocked off.  This gives rise to a one-dimensional streak centered on
each bright source which is aligned with the CCD row and column axes.
It ends up having $\sim 2$\% of the total number of events as the
original source\footnote{See
\url{http://cxc.harvard.edu/cal/Acis/Cal\_prods/xfer\_streak/index.html}.},
and can be used for both positional and spectral information.  Here we
make use of only the former.

\begin{figure}
\plotone{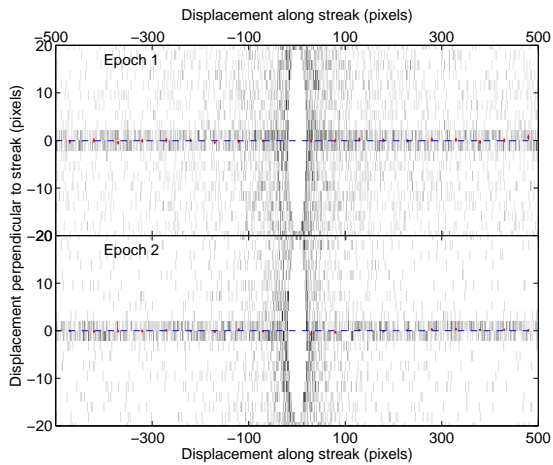}
\caption{Fits to the readout streak of \axp: epoch 1 (\textit{top})
  and epoch 2 (\textit{bottom}).  We rotated the event locations
  according to the \texttt{ROLL\_NOM} header keyword, ending up in a
  frame where the CCD readout streak was horizontal.  The grayscales
  are the binned images for each epoch; the bins appear rectangular in
  this plot because of the drastically different scales for the
  ordinates and abscissae.  We then constructed 50 pixel wide bins
  along the streak and computed the median event locations
  perpendicular to the streak, which we show as the points.  The
  dashed lines are the averages of the median positions.  We have
  excluded events within a radius of $20\,$pix around  \axp. }
\label{fig:streak}
\end{figure}

To use the streak, we took the event positions from the level-2 event
file.  We then rotated by the \texttt{ROLL\_NOM} header keyword,
putting them into a frame where the CCD rows and columns were vertical
and horizontal (similar to the \texttt{CHIP} frame).  The readout
streak, which is a  line in the sky ($x$,$y$) frame not aligned with
either axis, is now
horizontal.  We then stepped along the line, computing the median
event position in the vertical direction for bins of $50\,$pix width
in the horizontal direction: see Figure~\ref{fig:streak}.  For both
epochs 1 and 2, we found streak positions that were consistent to $\pm
0.02\,$pix with our best-fit positions of \axp\ from
Table~\ref{tab:refaxp}, albeit in one dimension. While not sufficient
to get a position for \axp\ alone, this provides additional
confirmation that our cross-correlation technique is accurate to the
claimed $0.1\,$pix.

\subsection{Frame Transformations}
\label{sec:tform}
We considered a number of ways to transform between the position
measurements at each stage.  To test this, we used 13 ACIS-I
observations\footnote{We used ObsIDs:
\dataset[ADS/Sa.CXO#obs/2232]{2232},
\dataset[ADS/Sa.CXO#obs/2233]{2233},
\dataset[ADS/Sa.CXO#obs/2234]{2234},
\dataset[ADS/Sa.CXO#obs/2421]{2421},
\dataset[ADS/Sa.CXO#obs/2423]{2423},
\dataset[ADS/Sa.CXO#obs/3293]{3293},
\dataset[ADS/Sa.CXO#obs/3294]{3294},
\dataset[ADS/Sa.CXO#obs/3388]{3388},
\dataset[ADS/Sa.CXO#obs/3389]{3389},
\dataset[ADS/Sa.CXO#obs/3390]{3390},
\dataset[ADS/Sa.CXO#obs/3391]{3391},
\dataset[ADS/Sa.CXO#obs/3408]{3408}, and
\dataset[ADS/Sa.CXO#obs/3409]{3409}.} of the Hubble Deep Field North
(HDF-N; \citealt{hbg+00} and subsequent papers).  The observations
were between 50~ks and 140~ks in depth, and detected $>100$~sources.
Note that the observational setup here was, if anything, more taxing
that what we used for our proper motion observations.  This was
because a number of the observations had a different aimpoint (off by
about half of an ACIS detector) and varying roll angles.  We
identified 38 sources that were common to each exposure (note that
this was not a completely exhaustive search, but comprised those that
we could identify easily) and computed transformations between each
pair of observations.  We used a number of different transformations
with different levels of complexity.
\begin{description}
\item[2-parameter]: A shift in $x$ and $y$.
\item[3-parameter]: A shift in $x$ and $y$, along with a rotation.
\item[4-parameter]: A shift in $x$ and $y$, along with a rotation and
  an overall scaling.
\item[6-parameter]: A general linear transformation, involving a shift
  and a separate rotation and scaling for each axis (e.g., see Eqn.~A5
  in \citealt*{kvka07}).
\end{description}
We were transforming between $x$ and $y$ values in the \texttt{SKY}
frame, so the nominal position angle of \chandra\ was already taken
out.  Any remaining rotation or scaling would account for imprecise
knowledge of the spacecraft's orientation and breathing of the ACIS-I
detector (changes in the plate-scale due to thermal variations), respectively.

\begin{figure}
\plotone{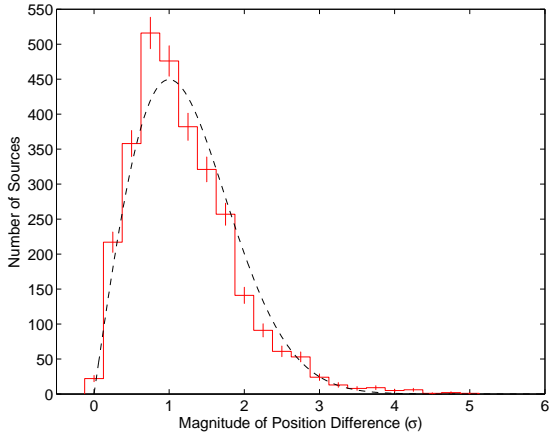}
\caption{Position differences for 2964 sources compared between
  different observations of the HDF-N (38 unique sources and 78 pairs
  of observations), normalized by position
  uncertainty (computed according to Eqn.~\ref{eqn:dposs}).
  The data are the histogram, and we also show the expected
  distribution with $N(r)\propto r e^{-r^2/2}$ for comparison.
}
\label{fig:pdifftform}
\end{figure}

Overall we found that the 3-parameter transformation provided a
significant decrease in $\chi^2$ compared to the 2-parameter
transformation, with a mean $\Delta \chi^2$ of 23.8 going from 74 to
73 DOF.  This is extremely significant, with an F-test probability of
$<10^{-5}$ that this is due to chance.  Therefore a non-zero rotation
is required between different observations, with a magnitude of
$\approx 0.05\degr$.  Comparing sets of three observations, the
rotations $\phi$ fitted between all of them are consistent, with
(e.g.,) $\phi(1\rightarrow2) + \phi(2\rightarrow3) \approx
\phi(1\rightarrow3)$, indicating that the rotations can be physically
meaningful.  With just the 3-parameter
transformation the reduced $\chi^2$ was consistent with 1.0, and the
distribution of position residuals was consistent with the expected
distribution (Fig.~\ref{fig:pdifftform}).  This indicates that our
uncertainty model is reasonable: tests using the ChaMPlane model gave
similar results, while those using the raw \texttt{wavdetect}
uncertainties had reduced $\chi^2$ larger than 2.
Beyond this, the more
complicated transformations were not warranted, with a median $\Delta
\chi^2$ of only 4.2 between 3 parameters and 6.

\section{Measuring Proper Motions}
\label{sec:pm}
\subsection{\sgr}
To actually measure the proper motion of \sgr, first we ran
\texttt{wavdetect} on both observations.  The parameters were the same as those
discussed in \S~\ref{sec:xray}.   We then identified sources common to
both epochs of the data, finding 27 sources in addition to  \sgr.
These sources were on all four of the ACIS-I CCDs.  It would be
preferable to have the \sgr\ centered on a single CCD, and to only use
sources on that detector.  That way we would not have to worry about
inaccuracies in the relative positions of the detectors or shifts in
their positions with time affecting our astrometry.  However, that
would severely restrict our field of view, and we felt that the gain
of additional reference sources outweighed the possibility of smaller
uncertainties for the detected sources.
The various tests and discussion in \S~\ref{sec:ref} seem to indicate
that using sources on different CCDs does not adversely affect our
astrometry, at least at the level considered here.

With the source detections, we then updated the position uncertainties
of all of the sources except  \sgr\ using Eqn.~\ref{eqn:dposs}.
Finally, we fit for the transformation between the two epochs of data,
using a 3-parameter fit (\S~\ref{sec:tform}).
The data gave a very good fit, with $\chi^2=37.7$ for 51 DOF (27
reference sources
and 3 parameters).  There were no significant outliers in the data,
which is not really surprising since we only included sources that we
could match initially between the epochs. The magnitude of the
residuals is $\sim 100\,\masyr$, which is far larger than
the proper motion one would expect from random field stars (i.e., at
1~kpc $100\,\masyr$ is $475\,\kms$, which is
much larger than the velocity dispersion of Galactic stars).
Therefore we can consider that we are measuring the motion of \sgr\
relative to reference stars that are at rest.  Similarly, corrections
for Galactic rotation and Solar motion are too small to be considered.

\begin{figure*}
\plottwo{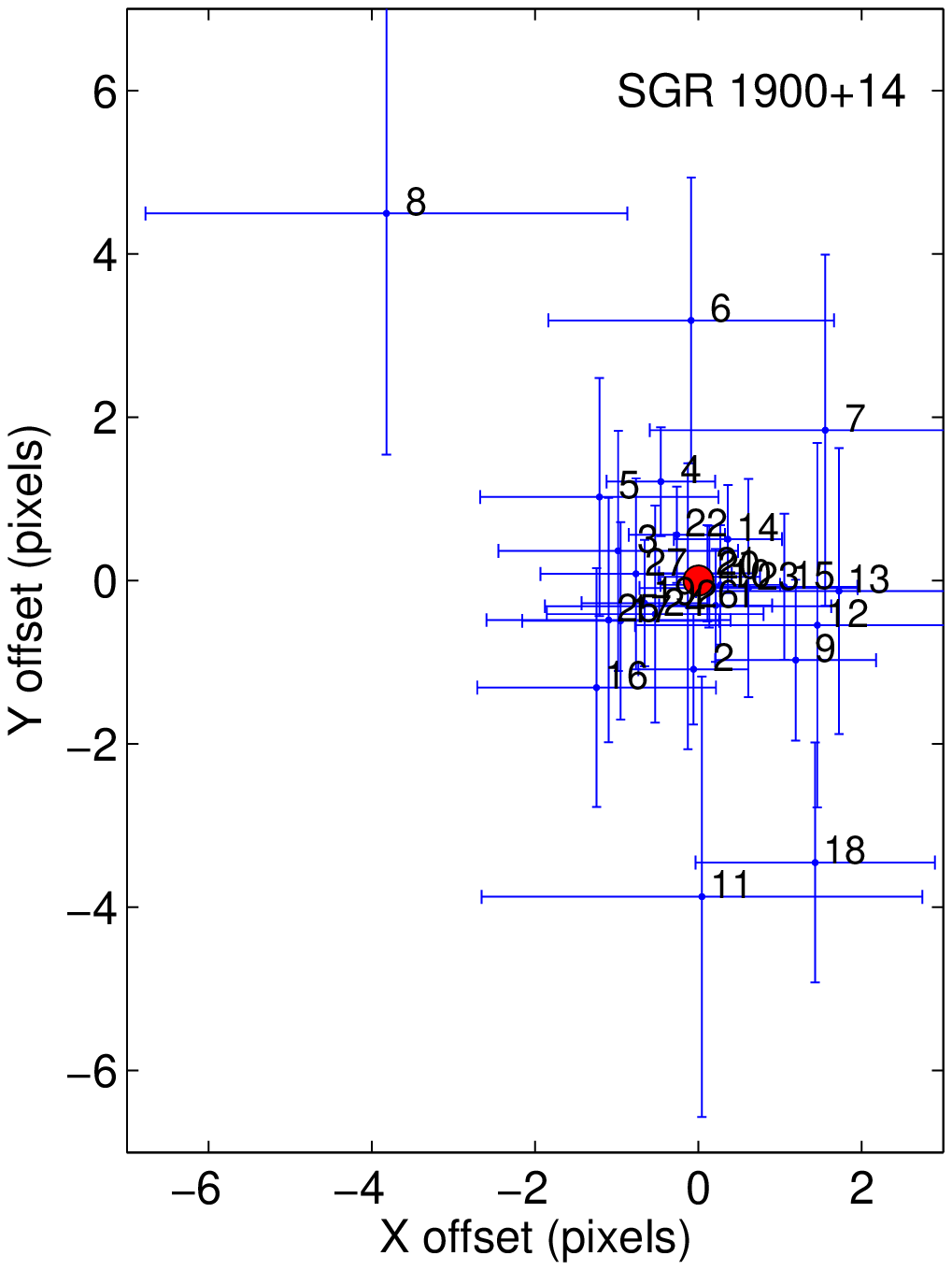}{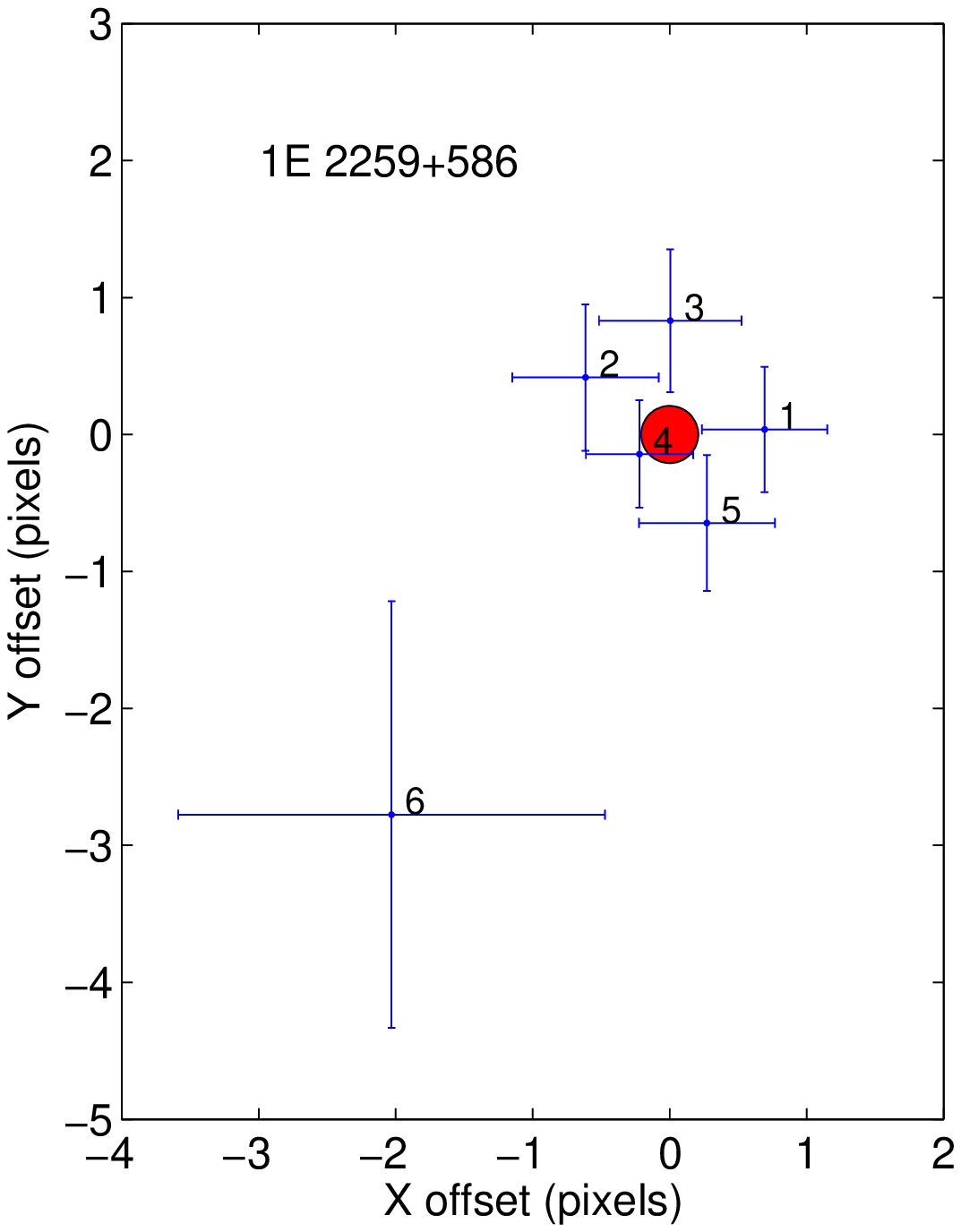}
\caption{Residual position differences between epochs 2 and 1 
  reference sources used to compute the transformation for \sgr\
  (\textit{Left}, with 27 reference sources) and \axp\
 (\textit{Right}, with 7 reference sources); the reference sources are
  labeled according to Tables~\ref{tab:ref} and \ref{tab:refaxp}.  The
  differences plotted are the residuals left after fitting for 
  2-parameter (i.e., shift only) transformations.  The filled circles at
  the centers show the uncertainty on the resulting transformations,
  which in this case is the uncertainty on the mean of the residuals.}
\label{fig:refresids}
\end{figure*}

The fitted rotation here was consistent with 0. We therefore found
that a 2-parameter transformation was sufficient (with virtually no
change in $\chi^2$), and with the smaller number of free parameters
the resulting position difference had smaller uncertainties.  While in
\S~\ref{sec:tform} we found that a rotation was necessary, here we do
not, although whether it is the simpler observational setup (both
observations had the same aimpoint, in contrast to \S~\ref{sec:tform})
or just the reduced accuracy from a smaller number of reference
sources that changed the situation we do not know.  We plot the
residual position differences of the reference sources in
Figure~\ref{fig:refresids}.

As a check on our transformation, we performed jackknife tests, where
we calculate the transformation for the data excluding a single
reference source (see, e.g., \citealt{et93}).  This tests whether or
not the
transformation depends overly on a single reference source,
and indeed we find that it does not.  We find that the means of the
jackknifed shifts are consistent with the value using all of the data,
and that the standard deviations are significantly less than our
estimated shift uncertainties.
We then performed a bootstrap test.  Here we created 100  data
sets,  each including a random subset  chosen with replacement from the 27
reference sources.  The resulting shifts are again distributed around
the proper values created from the actual data, and the standard
deviations are comparable to our uncertainties (they are actually
slightly smaller).  Based on these two tests, we see that our
transformation calculation is reasonably robust to choices of the
input data.

With the 3-parameter transformation, we find that the position of
\sgr\ is consistent between the two epochs, moving by
$0.27\pm0.27$~pixels, implying no measurable proper motion.  Note that
the bulk of this uncertainty comes from uncertainties in the fitted
transformation (uncertainties in each coordinate of the shift of
0.26~pixels, and an uncertainty in the rotation of $0.03\degr$), since
the individual position measurements of \sgr\ had uncertainties of
$<0.02$~pixels.  We also checked to see if the choice of
transformation (2, 3, 4, or 6 parameters), uncertainty model (ChaMP or
ChaMPlane), or level of processing (with or without ACIS
randomization) made a difference, and find that all of the results
were consistent.  With the 2-parameter transformation, which as
discussed above is sufficient, we find a very similar shift but
smaller residuals: $(\Delta x,\Delta
y)=(0.02\pm0.19,0.26\pm0.19)$~pixels, or a magnitude
$0.27\pm0.19$~pixels for the 2-parameter fit.  Again, the
uncertainties are dominated by the 0.18~pixel uncertainties in each
coordinate of the shift.  We give the detailed fit results in
Table~\ref{tab:pm}.

The interval between the observations was $\Delta t=1813.3\,\mbox{days}=4.96\mbox{ yr}$, so we have a proper motion in pixels
$(\mu_x,\mu_y)=(2.4\pm18.4,26.3\pm18.4)\,\masyr$.  The
celestial proper motion is $(\mu_\alpha,\mu_\delta)=0\farcs492(-\mu_x,\mu_y)$,
which is $(\mu_\alpha,\mu_\delta)=(-2.4\pm18.4,26.3\pm18.4)\,\masyr$. If we actually want to limit the speed of \sgr, we can look
at the distribution of the magnitude $\mu=\sqrt{\mu_x^2+\mu_y^2}$.
Our measurements were $\mu_x=\hat\mu_x\pm\sigma_{\mu,x}$ and
$\mu_y=\hat\mu_y\pm\sigma_{\mu,y}$, although
$\sigma_{\mu,x}=\sigma_{\mu,y}=\sigma_\mu$.  If we look at the
magnitude of the proper motion $\mu=\hat\mu \pm \sigma_\mu$, where
$\hat \mu=\sqrt{\hat \mu_x^2+\hat \mu_y^2}$, the probability density
function for $\mu$ is:
\begin{equation}
f_{\mu}(\mu)=\left(\frac{\mu}{\sigma_\mu^2}\right) I_0\left(
\frac{\mu \hat{\mu}}{\sigma_\mu^2}\right) e^{-(\mu^2+\hat
  \mu^2)/2\sigma_\mu^2},
\label{eqn:pm}
\end{equation}
where $I_0(x)$ is the modified Bessel function of the first kind
\citep[][p.\ 140]{papoulis91}.  
From numerical integration, we find that the mean, mode, and median
are all close to each other.  For \sgr, the mean is $33\,\masyr$ (vs.\ $\hat \mu=26\,\masyr$), and we find a 90\%
upper limit to $\mu$ to be $54\,\masyr$, or a transverse
velocity $v_{\perp} < 1300d_5\,\kms$.

\begin{deluxetable*}{l c c c c c}
\tablecaption{Proper Motion Measurements for \sgr\ and \axp\label{tab:pm}}
\tablewidth{0pt}
\tabletypesize{\scriptsize}
\tablehead{
\colhead{Parameter} & \mc{2}{c}{\sgr} && \mc{2}{c}{\axp} \\
\cline{2-3} \cline{5-6}
 & \colhead{2-parameter} & \colhead{3-parameter} &&\colhead{2-parameter} & \colhead{3-parameter} \\
}
\startdata
Transform $\chi^2/{\rm DOF}$ & 37.8/52 & 37.7/51 && 14.1/10 & 13.1/9\\
$(\Delta x, \Delta y)$ (pixels)\tablenotemark{a} & $(0.02\pm0.19, 0.26\pm0.19)$& $(0.02\pm0.27, 0.26\pm0.27)$ && $(-0.14\pm0.25,
-0.45\pm0.25)$& $(-0.15\pm0.33, -0.39\pm0.34)$\\
$(\mu_x, \mu_y)$ $(\masyr)$\tablenotemark{b}  &  $(2\pm18, 26\pm18)$ & $(2\pm26, 26\pm26)$ &&
$(-11\pm20, -35\pm20)$&$(-11\pm26,-31\pm26)$\\
$\hat \mu$ $(\masyr)$\tablenotemark{c}& 26 & 26&& 36 & 32\\
$\langle\mu\rangle$ $(\masyr)$\tablenotemark{d} & 33 & 39&& 42 & 44\\
$\mu_{\rm 90}$ $(\masyr)$\tablenotemark{e} & 54 & 69& & 65 & 73\\
$\langle v_{\perp}\rangle$ $({\rm km}\,{\rm s}^{-1})$\tablenotemark{f}
& 780$d_5$ & 930$d_5$ && 600$d_{3}$& 630$d_{3}$\\
$v_{\perp,90}$ $({\rm km}\,{\rm s}^{-1})$\tablenotemark{g} & 1300$d_5$ & 1600$d_5$&&930$d_{3}$& 1000$d_{3}$\\
\enddata
\tablenotetext{a}{Shift  in pixels between epochs.}
\tablenotetext{b}{Proper motion  in the $(x_{\rm sky},y_{\rm sky})$ frame between epochs.}
\tablenotetext{c}{Magnitude of the proper motion, computed
  from $\sqrt{\mu_x^2+\mu_y^2}$.}
\tablenotetext{d}{Mean of the magnitude of the proper motion, from
  the distribution in Eqn.~\ref{eqn:pm}.}
\tablenotetext{e}{90\% upper limit to the magnitude of the proper motion, from
  the distribution in Eqn.~\ref{eqn:pm}.}
\tablenotetext{f}{Transverse velocity corresponding to
  $\langle\mu\rangle$, the mean of the magnitude of the proper motion, assuming a distance of $5d_5\,$kpc (\sgr) or
  $3d_{3}\,$kpc (\axp).}
\tablenotetext{g}{Transverse velocity corresponding to the $\mu_{90}$,
  the 90\% upper
  limit on the proper motion, assuming a distance of $5d_5\,$kpc (\sgr) or
  $3d_{3}\,$kpc (\axp).}
\end{deluxetable*}

\begin{figure*}
\plottwo{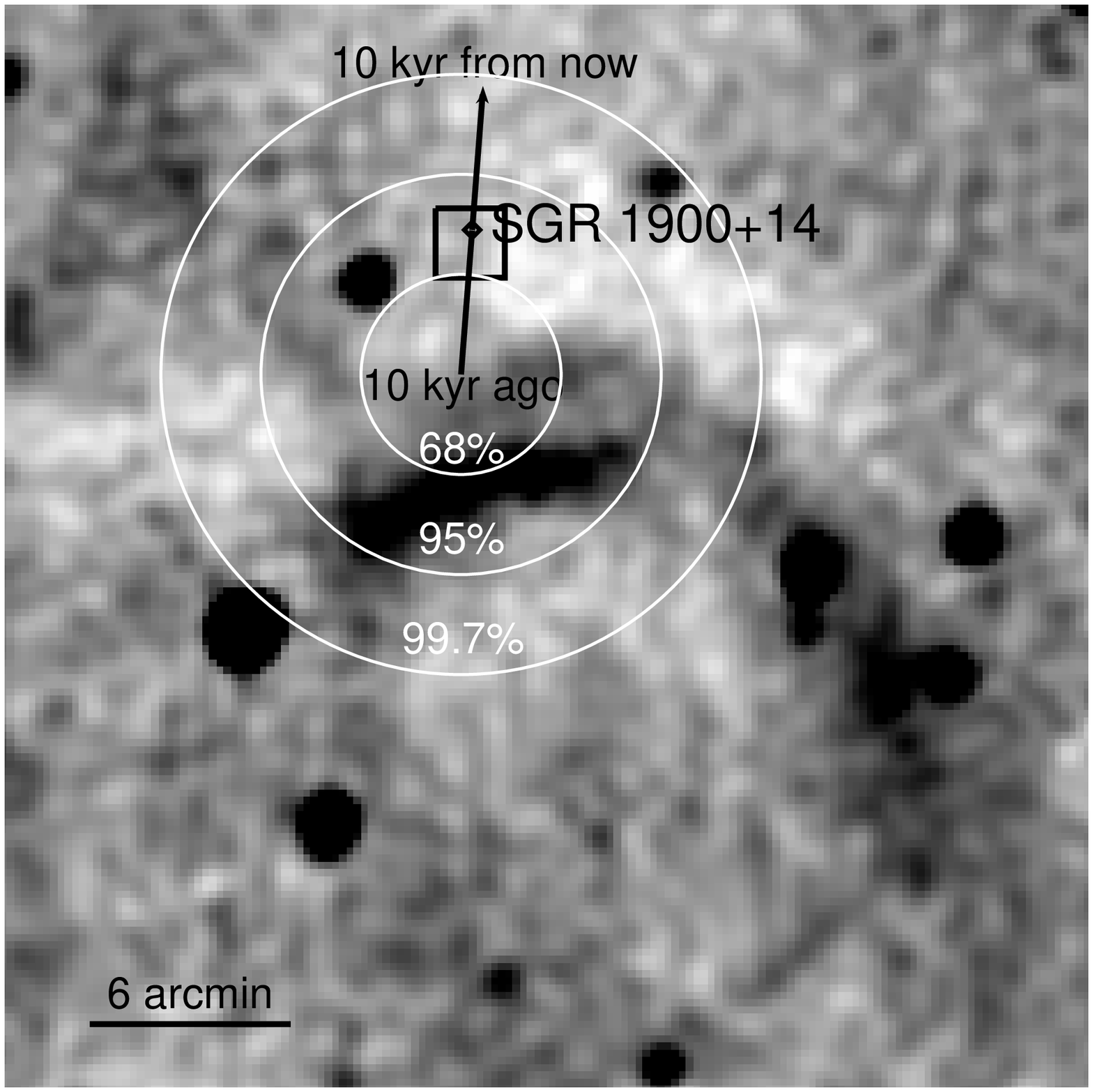}{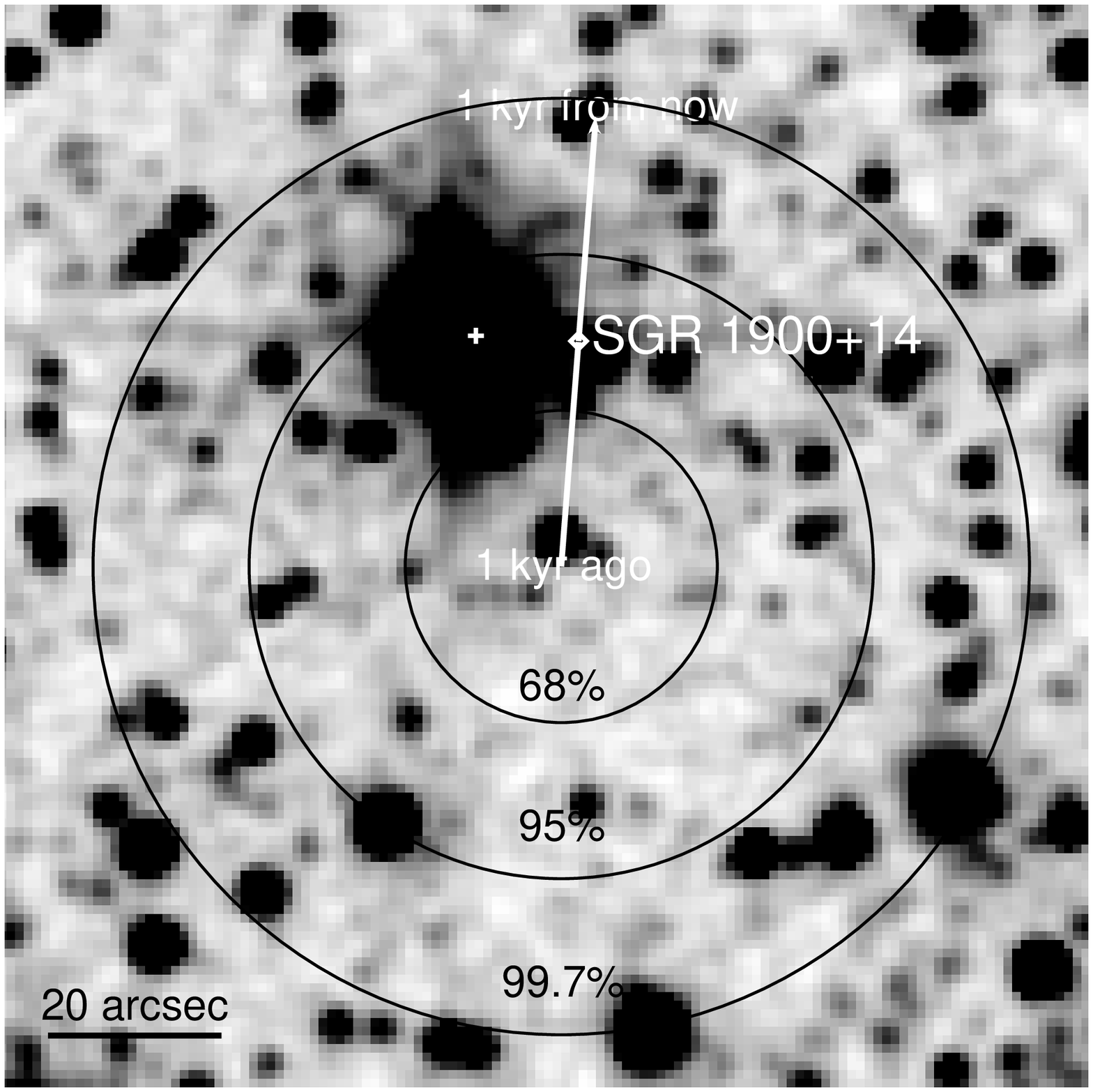}
\caption{\textit{Left:} Radio image of the field around \sgr\
  (\citealt*{fkb99}; marked with a diamond), showing \snr\ (to the
  south of \sgr).  The image is a 332$\,$MHz
  image from \citet{kkfvk02}.  The arrow shows the
  distance that \sgr\ has moved, starting 10$\,$kyr ago and projecting
  the same time into the future  using our
  best-fit proper motion from Table~\ref{tab:pm}.  We also show  1-, 2-, and
  3-$\sigma$ error circles (68\%, 95\%, and 99.7\% confidence)
  on the position 10$\,$kyr ago.  {Right:} the
  same but plotted on a near-infrared ($2.2\,\um$) image from the Two
  Micron All Sky Survey (2MASS; \citealt{2mass}).  The position of the cluster found by
  \citet[][]{vhl+00} is marked with a cross.  The scale of this
  image is reduced by a factor of $\sim 10$ from the radio image (the
  box on the radio image is the field shown in the near-IR image), as the
  proper motion vector now only shows the distance that \sgr\ has moved/will
  move over 1$\,$kyr. There are scale bars in the lower left
  corners, and the images have north up and east to the left.}
\label{fig:radiosgr}
\end{figure*}

\subsection{\axp}
\label{sec:axp}
We followed the same procedure as for \sgr\ above to find the
reference sources for \axp.  Here, though, we could only identify 6 sources
common to both epochs (a seventh source was identified, but it is
right near the gap between detectors and the positions disagreed by
$\approx 3\,\sigma$ between the two epochs).  While formally enough to
determine a robust transformation, this is fewer than we might like to
have.  One problem here is that we are using ACIS-S, which has a
smaller field of view.  Here \axp\ is close to the center of a CCD
(ACIS-S3), and we used sources from both ACIS-S3 and ACIS-S2.  We
chose to do the second epoch observation with ACIS-S as well, to best
reproduce the first.  There were some slight problems, though.  As the
roll angle changed by $121\degr$, the position of ACIS-S2 changed
relative to the aim-point.  Therefore, some of the reference sources
detected in the first epoch were missed entirely by the detector in
the second epoch, and vice versa.  This also explains why some of the
reference sources have such drastically different count-rates between
the two epochs: some of them move from the front-illuminated (and
hence less sensitive) ACIS-S2 to the back-illuminated ACIS-S3,
although in other cases the difference in count-rate is intrinsic, and
may reflect flare stars or other variable sources.

We computed 2- and 3-parameter transformations.  As with \sgr, the
3-parameter transformation did not provide any improvement.  With the
2-parameter transformation, we find the position residuals in
Figure~\ref{fig:refresids}.  We also repeated the jackknife and
bootstrap tests (\S~\ref{sec:pm}).  The residuals in
Figure~\ref{fig:refresids} show an outlier, but it is relatively minor
and the jackknifed transformations are consistent with the mean.

With the transformations from the six reference sources, we again find
only an upper limit to a displacement between the two epochs:
$\mu<65\,\masyr$ at 90\% confidence (translating to $v_{\perp} < 930
d_{3}\,\kms$), with additional details given in Table~\ref{tab:pm}.
Unlike \sgr, where the reduced $\chi^2$ for the 2-parameter
transformation was less than 1.0, here it is 1.4.  This appears
slightly worrisome, and could indicate that we have underestimated our
uncertainties, which would make the inferred proper motion less
significant.  However, the majority of the problem is caused by a
single source, \#6 in Table~\ref{tab:refaxp}.  This source deviates by
$\approx 1.5\,\sigma$, and without it the reduced $\chi^2$ becomes
1.1.  Including source \#6, it is still not impossible to get such a
large reduced $\chi^2$ even with a proper description of the data: it
should happen about 17\% of the time, so our result is not really
surprising (also recall that our tests with the HDF-N observations
gave a reduced $\chi^2$ of 1.33 for 1450 DOF with the ChaMP model).
Overall, the residuals for the remaining sources are reasonably well
behaved, and given the tests for \sgr\ and those in \S~\ref{sec:ref},
we believe that our uncertainties are on the whole reliable.

\begin{figure}
\plotone{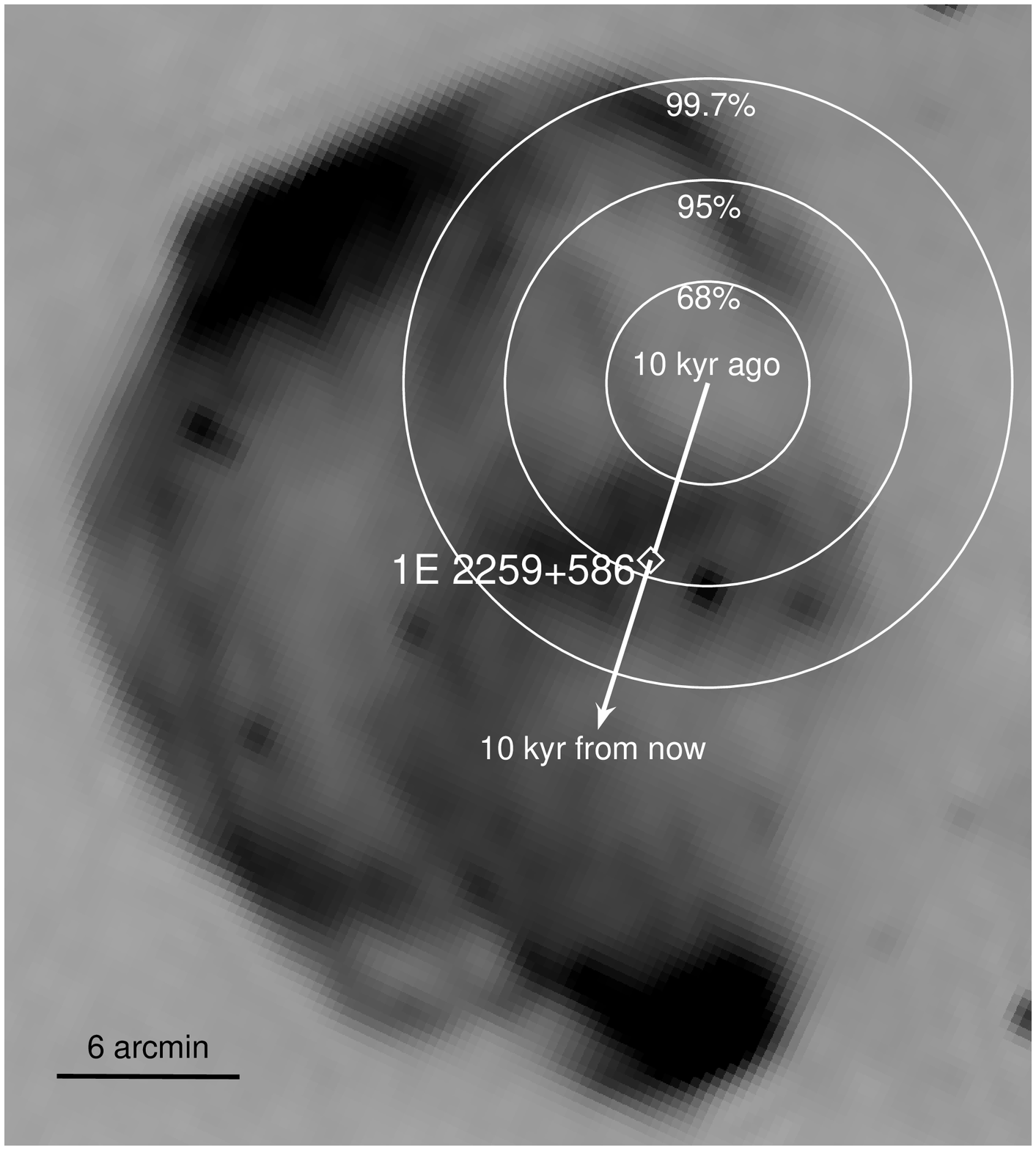}
\caption{Radio image of the SNR CTB~109 at 1.4$\,$GHz, with the
  position of \axp\ marked.  The image is from the Canadian Galactic
  Plane Survey (CGPS; \citealt{tgp+03}).  The diamond shows the
  current position of \axp\ \citep{htvk+01}.  The arrow shows the
  distance that \axp\ has moved, starting 10$\,$kyr ago and projecting
  the same time into the future  using our
  best-fit proper motion from Table~\ref{tab:pm}.  We also show  1-, 2-, and
  3-$\sigma$ error circles (68\%, 95\%, and 99.7\% confidence)
  on the position 10$\,$kyr ago.  The scale bar in the lower left
  indicates $6\arcmin$, and the image has north up and east to the left.}
\label{fig:radioaxp}
\end{figure}

\subsection{Prospects for Improvement}
Our result here is formally a non-detection of the proper motions of
\sgr\ and \axp.  However, the direction and magnitude of \sgr\ are
suggestive, pointing as they do to a nearby SNR (see
\S~\ref{sec:disc}).  It would therefore be interesting to confirm or
refute our measurements with newer data.  Beyond this, one must always
view 2-point proper motions with suspicion, as a line can always be
fit to two points, and only with the inclusion of a third observation
does one begin to gain confidence in the result.  The biggest limiting
factor is astrometry of the reference sources (this is similar to what
was found by \citealt{mphs07,mph+07}): increasing the number or
accuracy of those measurements would be ideal, but that would have to
be done for all of the observations.  Just improving the newest one
would have little impact.  If we assume similar observations to those
presented here, we can achieve frame matching to a precision of
$\approx 0.2$~pixels between a pair of observations.  To measure a
proper motion of the order of $\sim 26\,\masyr$, we would need to wait
$>10$~yrs from the first observation for a statistically significant
($>3\,\sigma$) measurement.  This is only considering the first and
last data-sets, though: by analyzing all three observations together
we can establish a more robust reference frame and confirm the
measurement of displacement.  Regardless, we would need to wait until
2010 or so before this will be possible.  Observations at other
wavelengths, possible for \axp\ at least, should be able to achieve
higher precision \citep[][P.~B.~Cameron 2008, pers. comm.]{cbk08}.

Another approach would be to use the High Resolution Camera (HRC;
\citealt{m+97}) on \chandra.  The pixel scale is almost factor of four
smaller than that of ACIS, so the HRC has the potential for better
spatial resolution.  However, the effective resolution is $\approx
0\farcs5$, not that different from ACIS.  It also does not suffer from
pileup, so bright sources can be observed without penalty.  The main
drawback is that because of lower overall efficiency few, if any,
reference sources are detected: \citet{hb06} and \citet{wp07} found
only two and three sources, respectively.  The accuracy of our present
work is limited by the registration of the separate observations which
depends on the reference sources.  This was true even for \axp, where
pileup limited the centroiding accuracy.  Even so, \citet{wp07} found
a significant proper motion of $165\pm25\,\masyr$ for RX~J0822$-$4300
using three HRC observations (also see \citealt{hb06} for an earlier
but consistent result).

\section{Discussion and Conclusions}
\label{sec:disc}
Our proper motion limit can be used to limit the age of \sgr/\snr,
assuming an association. From the 327-MHz radio data from
\citet{kkfvk02}, we represent \snr\ as a ring centered at (J2000):
$\alpha=19^{\rm h}06^{\rm m}59\fs4$,
$\delta=+09\degr03\arcmin05\arcsec$, and with a radius of $\approx
9.5\arcmin$ (see Fig.~\ref{fig:radiosgr}).  \sgr\ would have needed $>16$~kyr to traverse the
separation of $15\arcmin$, using the 90\% confidence upper limit on
the proper motion.  This is not a remarkably large age for a magnetar \citep{gsgv01,wt06}.  Suggestively, the
marginal proper motion that we find is directed to the north, and
\sgr\ sits beyond the northern rim of \snr.  However, our proper
motion is so uncertain in both direction and magnitude that 
 an association between \sgr\ and the cluster
found by \citet{vhl+00} is also consistent with the data.  

For \axp, we find an upper limit to the proper motion.
The SNR CTB~109 (see Fig.~\ref{fig:radioaxp}), with which \axp\ is
generally taken to be associated, has a somewhat complicated
morphology \citep{kuy02} with what may be multiple shells and seems to
be interacting with a giant molecular cloud to the west
\citep{tfi+90}.  The site of the explosion is not particularly well
defined, given the different shells and the impeded expansion to the
west.  However, there does not appear to be any suggestion that \axp\
is displaced to the south relative to the explosion center, as our
proper motion limit might suggest.  The magnitude of the proper motion limit is not huge: for
an age of $1000t_{3}\,{\rm yr}$, \axp\ would have moved $1
t_{3}\,$arcmin, so it could have still had an origin within the
current $\approx 30\arcmin$-diameter shell for ages up to $t_3\sim
10$.  Therefore our data do not greatly constrain the association
between \axp\ and CTB~109 (generally assumed to be secure).

Taken together, we can try to use the velocities we measure/limit to
understand if magnetars have a velocity distribution comparable to
that of the radio pulsar population.  The proper motion of
XTE~J1810$-$197 \citep{hcb+07} and the velocity limits inferred by
\citet{gsgv01} are less than $500\,\kms$.  For the magnetars
considered here, though our results are less constraining.  This is
made worse by the very uncertain distances to these objects.  If we
accept the association between \sgr\ and the massive star cluster of
\citet[][also see \citealt{wrrd+08}]{vhl+00}, then we must also accept
a distance to \sgr\ of $\approx 15\,$kpc.  This means that our 90\%
upper limit on the transverse velocity is 3800$\,{\rm km}\,{\rm
s}^{-1}$, which encompasses the extreme range of velocities posited
for the magnetars \citep[e.g.,][]{dt92,td93,td95}, but the fact of the
association itself would imply a small proper motion on the order of
$\sim 5\,\masyr$.  For \axp, our transverse velocity limit is
moderately constraining (although with the larger distance of
\citealt{dvk06} it increases to almost 2500$\,{\rm km}\,{\rm
s}^{-1}$).

Future astrometric observations will help answer some of these
questions.  We now understand in detail what the limiting factors are,
and a third epoch of data around 2010 should allow definitive measurement of the
proper motions presented here.  However, the velocity situation will
 remain more uncertain, owing to the factor of 3
uncertainty in the distance to \sgr\ as well as unknown projection effects.

\acknowledgements Support for this work was provided by the National
Aeronautics and Space Administration through Chandra award GO6-7066X.
Partial support for DLK was also provided by NASA through Hubble
Fellowship grant \#01207.01-A awarded by the Space Telescope Science
Institute, which is operated by the Association of Universities for
Research in Astronomy, Inc., for NASA, under contract NAS 5-26555.
B.M.G. acknowledges the support of a Federation Fellowship from the
Australian Research Council through grant FF0561298.  Support for SC
was provided by the University of Sydney Postdoctoral Fellowship
program. POS also acknowledges support from NASA contract
NAS8-201039073.  The research presented in this paper has used data
from the Canadian Galactic Plane Survey, a Canadian project with
international partners, supported by the Natural Sciences and
Engineering Research Council.


\begin{thebibliography}{}

\bibitem[{Cameron} {et~al.}(2008){Cameron}, {Britton}, \& {Kulkarni}]{cbk08}
{Cameron}, P.~B., {Britton}, M.~C., \& {Kulkarni}, S.~R. 2008, \aj, submitted,  arXiv:0805.2153

\bibitem[{Carter} {et~al.}(2003){Carter}, {Karovska}, {Jerius}, {Glotfelty},  \& {Beikman}]{ckj+03}
{Carter}, C., {Karovska}, M., {Jerius}, D., {Glotfelty}, K., \& {Beikman}, S.  2003, in Astronomical Society of the Pacific Conference Series, Vol. 295,  Astronomical Data Analysis Software and Systems XII, ed. H.~E. {Payne}, R.~I.  {Jedrzejewski}, \& R.~N. {Hook} (San Francisco: ASP), 477

\bibitem[{Chatterjee} {et~al.}(2005){Chatterjee}, {Vlemmings}, {Brisken},  {Lazio}, {Cordes}, {Goss}, {Thorsett}, {Fomalont}, {Lyne}, \&  {Kramer}]{cvb+05}
{Chatterjee}, S., {et al.} 2005, \apjl, 630, L61

\bibitem[{Duncan} \& {Thompson}(1992){Duncan} \& {Thompson}]{dt92}
{Duncan}, R.~C. \& {Thompson}, C. 1992, \apjl, 392, L9

\bibitem[{Durant} \& {van Kerkwijk}(2006){Durant} \& {van Kerkwijk}]{dvk06}
{Durant}, M. \& {van Kerkwijk}, M.~H. 2006, \apj, 650, 1070

\bibitem[{Efron} \& {Tibshirani}(1993){Efron} \& {Tibshirani}]{et93}
{Efron}, B. \& {Tibshirani}, R.~J. 1993, An Introduction to the Bootstrap  (London: Chapman and Hall)

\bibitem[{Fahlman} \& {Gregory}(1981){Fahlman} \& {Gregory}]{fg81}
{Fahlman}, G.~G. \& {Gregory}, P.~C. 1981, \nat, 293, 202

\bibitem[{Frail} {et~al.}(1999){Frail}, {Kulkarni}, \& {Bloom}]{fkb99}
{Frail}, D.~A., {Kulkarni}, S.~R., \& {Bloom}, J.~S. 1999, \nat, 398, 127

\bibitem[{Freeman} {et~al.}(2002){Freeman}, {Kashyap}, {Rosner}, \&  {Lamb}]{fkrl02}
{Freeman}, P.~E., {Kashyap}, V., {Rosner}, R., \& {Lamb}, D.~Q. 2002, \apjs,  138, 185

\bibitem[{Gaensler} {et~al.}(2005){Gaensler}, {McClure-Griffiths}, {Oey},  {Haverkorn}, {Dickey}, \& {Green}]{gmo+05}
{Gaensler}, B.~M., {McClure-Griffiths}, N.~M., {Oey}, M.~S., {Haverkorn}, M.,  {Dickey}, J.~M., \& {Green}, A.~J. 2005, \apjl, 620, L95

\bibitem[{Gaensler} {et~al.}(2001){Gaensler}, {Slane}, {Gotthelf}, \&  {Vasisht}]{gsgv01}
{Gaensler}, B.~M., {Slane}, P.~O., {Gotthelf}, E.~V., \& {Vasisht}, G. 2001,  \apj, 559, 963

\bibitem[{Garmire} {et~al.}(2003){Garmire}, {Bautz}, {Ford}, {Nousek}, \&  {Ricker}]{gbf+03}
{Garmire}, G.~P., {Bautz}, M.~W., {Ford}, P.~G., {Nousek}, J.~A., \& {Ricker},  G.~R. 2003, \procspie, 4851, 28

\bibitem[{Gavriil} {et~al.}(2008){Gavriil}, {Gonzalez}, {Gotthelf}, {Kaspi},  {Livingstone}, \& {Woods}]{ggg+08}
{Gavriil}, F.~P., {Gonzalez}, M.~E., {Gotthelf}, E.~V., {Kaspi}, V.~M.,  {Livingstone}, M.~A., \& {Woods}, P.~M. 2008, Science, 319, 1802

\bibitem[{Gotthelf} {et~al.}(2004){Gotthelf}, {Halpern}, {Buxton}, \&  {Bailyn}]{ghbb04}
{Gotthelf}, E.~V., {Halpern}, J.~P., {Buxton}, M., \& {Bailyn}, C. 2004, \apj,  605, 368

\bibitem[{Halpern} {et~al.}(2005){Halpern}, {Gotthelf}, {Becker}, {Helfand},  \& {White}]{hgb+05}
{Halpern}, J.~P., {Gotthelf}, E.~V., {Becker}, R.~H., {Helfand}, D.~J., \&  {White}, R.~L. 2005, \apjl, 632, L29

\bibitem[{Helfand} {et~al.}(2007){Helfand}, {Chatterjee}, {Brisken}, {Camilo},  {Reynolds}, {van Kerkwijk}, {Halpern}, \& {Ransom}]{hcb+07}
{Helfand}, D.~J., {Chatterjee}, S., {Brisken}, W.~F., {Camilo}, F., {Reynolds},  J., {van Kerkwijk}, M.~H., {Halpern}, J.~P., \& {Ransom}, S.~M. 2007, \apj,  662, 1198

\bibitem[{Hobbs} {et~al.}(2005){Hobbs}, {Lorimer}, {Lyne}, \&  {Kramer}]{hllk05}
{Hobbs}, G., {Lorimer}, D.~R., {Lyne}, A.~G., \& {Kramer}, M. 2005, \mnras,  360, 974

\bibitem[{Hong} {et~al.}(2005){Hong}, {van den Berg}, {Schlegel}, {Grindlay},  {Koenig}, {Laycock}, \& {Zhao}]{hvdbs+05}
{Hong}, J., {van den Berg}, M., {Schlegel}, E.~M., {Grindlay}, J.~E., {Koenig},  X., {Laycock}, S., \& {Zhao}, P. 2005, \apj, 635, 907

\bibitem[{Hornschemeier} {et~al.}(2000){Hornschemeier}, {Brandt}, {Garmire},  {Schneider}, {Broos}, {Townsley}, {Bautz}, {Burrows}, {Chartas}, {Feigelson},  {Griffiths}, {Lumb}, {Nousek}, \& {Sargent}]{hbg+00}
{Hornschemeier}, A.~E., {et al.} 2000, \apj, 541, 49

\bibitem[{Hui} \& {Becker}(2006){Hui} \& {Becker}]{hb06}
{Hui}, C.~Y. \& {Becker}, W. 2006, \aap, 457, L33

\bibitem[{Hulleman} {et~al.}(2001){Hulleman}, {Tennant}, {van Kerkwijk},  {Kulkarni}, {Kouveliotou}, \& {Patel}]{htvk+01}
{Hulleman}, F., {Tennant}, A.~F., {van Kerkwijk}, M.~H., {Kulkarni}, S.~R.,  {Kouveliotou}, C., \& {Patel}, S.~K. 2001, \apjl, 563, L49

\bibitem[{Hurley} {et~al.}(1999a){Hurley}, {Cline}, {Mazets},  {Barthelmy}, {Butterworth}, {Marshall}, {Palmer}, {Aptekar}, {Golenetskii},  {Il'Inskii}, {Frederiks}, {McTiernan}, {Gold}, \& {Trombka}]{hcm+99}
{Hurley}, K., {et al.} 1999a, \nat, 397, 41

\bibitem[{Hurley} {et~al.}(1999b){Hurley}, {Kouveliotou},  {Woods}, {Cline}, {Butterworth}, {Mazets}, {Golenetskii}, \&  {Frederics}]{hkw+99b}
{Hurley}, K., {Kouveliotou}, C., {Woods}, P., {Cline}, T., {Butterworth}, P.,  {Mazets}, E., {Golenetskii}, S., \& {Frederics}, D. 1999b,  \apjl, 510, L107

\bibitem[{Hurley} {et~al.}(1999c){Hurley}, {Li}, {Kouveliotou},  {Murakami}, {Ando}, {Strohmayer}, {van Paradijs}, {Vrba}, {Luginbuhl},  {Yoshida}, \& {Smith}]{hlk+99}
{Hurley}, K., {et al.} 1999c, \apjl, 510, L111

\bibitem[{Hurley} {et~al.}(1996){Hurley}, {Li}, {Vrba}, {Luginbuhl},  {Hartmann}, {Kouveliotou}, {Meegan}, {Fishman}, {Kulkarni}, {Frail},  {Bowyer}, \& {Lampton}]{hlv+96}
{Hurley}, K., {et al.} 1996, \apjl, 463, L13

\bibitem[{Kaplan} {et~al.}(2002){Kaplan}, {Kulkarni}, {Frail}, \& {van  Kerkwijk}]{kkfvk02}
{Kaplan}, D.~L., {Kulkarni}, S.~R., {Frail}, D.~A., \& {van Kerkwijk}, M.~H.  2002, \apj, 566, 378

\bibitem[{Kaplan} {et~al.}(2007){Kaplan}, {van Kerkwijk}, \&  {Anderson}]{kvka07}
{Kaplan}, D.~L., {van Kerkwijk}, M.~H., \& {Anderson}, J. 2007, \apj, 660, 1428

\bibitem[{Kaspi} \& {McLaughlin}(2005){Kaspi} \& {McLaughlin}]{km05}
{Kaspi}, V.~M. \& {McLaughlin}, M.~A. 2005, \apjl, 618, L41

\bibitem[{Kasumov} {et~al.}(2006){Kasumov}, {Allakhverdiev}, \&  {Asvarov}]{kaa06}
{Kasumov}, F.~K., {Allakhverdiev}, A.~O., \& {Asvarov}, A.~I. 2006, Astronomy  Letters, 32, 308

\bibitem[{Kim} {et~al.}(2007){Kim}, {Kim}, {Wilkes}, {Green}, {Kim},  {Anderson}, {Barkhouse}, {Evans}, {Ivezi{\'c}}, {Karovska}, {Kashyap}, {Lee},  {Maksym}, {Mossman}, {Silverman}, \& {Tananbaum}]{kkw+07}
{Kim}, M., {et al.} 2007, \apjs, 169, 401

\bibitem[{Kothes} {et~al.}(2002){Kothes}, {Uyaniker}, \& {Yar}]{kuy02}
{Kothes}, R., {Uyaniker}, B., \& {Yar}, A. 2002, \apj, 576, 169

\bibitem[{Kouveliotou} {et~al.}(1994){Kouveliotou}, {Fishman}, {Meegan},  {Paciesas}, {van Paradijs}, {Norris}, {Preece}, {Briggs}, {Horack},  {Pendleton}, \& {Green}]{kfm+94}
{Kouveliotou}, C., {et al.} 1994, \nat, 368, 125

\bibitem[{Kouveliotou} {et~al.}(1999){Kouveliotou}, {Strohmayer}, {Hurley},  {van~Paradijs}, {Finger}, {Dieters}, {Woods}, {Thompson}, \&  {Duncan}]{ksh+99}
{Kouveliotou}, C., {et al.} 1999, \apjl, 510, L115

\bibitem[{Kulkarni} {et~al.}(2003){Kulkarni}, {Kaplan}, {Marshall}, {Frail},  {Murakami}, \& {Yonetoku}]{kkm+03}
{Kulkarni}, S.~R., {Kaplan}, D.~L., {Marshall}, H.~L., {Frail}, D.~A.,  {Murakami}, T., \& {Yonetoku}, D. 2003, \apj, 585, 948

\bibitem[{Lai}(2001){Lai}]{lai01b}
{Lai}, D. 2001, in Lecture Notes in Physics, Vol. 578, Physics of Neutron Star  Interiors, ed. D.~{Blaschke}, N.~K. {Glendenning}, \& A.~{Sedrakian} (Berlin:  Springer Verlag), 424

\bibitem[{Lampton} {et~al.}(1976){Lampton}, {Margon}, \& {Bowyer}]{lmb76}
{Lampton}, M., {Margon}, B., \& {Bowyer}, S. 1976, \apj, 208, 177

\bibitem[{Lin} \& {Zhang}(2004){Lin} \& {Zhang}]{lz04}
{Lin}, J.~R. \& {Zhang}, S.~N. 2004, \apjl, 615, L133

\bibitem[{Lorimer} \& {Xilouris}(2000){Lorimer} \& {Xilouris}]{lx00}
{Lorimer}, D.~R. \& {Xilouris}, K.~M. 2000, \apj, 545, 385

\bibitem[{Mazets} {et~al.}(1979){Mazets}, {Golenetskij}, \& {Guryan}]{mgg79}
{Mazets}, E.~P., {Golenetskij}, S.~V., \& {Guryan}, Y.~A. 1979, Soviet  Astronomy Letters, 5, 343+

\bibitem[{Mereghetti} {et~al.}(2006){Mereghetti}, {Esposito}, {Tiengo},  {Zane}, {Turolla}, {Stella}, {Israel}, {G{\"o}tz}, \& {Feroci}]{met+06}
{Mereghetti}, S., {et al.} 2006, \apj,  653, 1423

\bibitem[{Motch} {et~al.}(2007a){Motch}, {Pires}, {Haberl}, \&  {Schwope}]{mphs07}
{Motch}, C., {Pires}, A.~M., {Haberl}, F., \& {Schwope}, A. 2007a,  \apss, 69

\bibitem[{Motch} {et~al.}(2007b){Motch}, {Pires}, {Haberl},  {Schwope}, \& {Zavlin}]{mph+07}
{Motch}, C., {Pires}, A.~M., {Haberl}, F., {Schwope}, A., \& {Zavlin}, V.~E.  2007b, in {40 Years of Pulsars: Millisecond Pulsars, Magnetars  and More}, Vol. 712, arXiv:0712.0342

\bibitem[{Muno} {et~al.}(2006){Muno}, {Clark}, {Crowther}, {Dougherty}, {de  Grijs}, {Law}, {McMillan}, {Morris}, {Negueruela}, {Pooley}, {Portegies  Zwart}, \& {Yusef-Zadeh}]{mcc+06}
{Muno}, M.~P., {et al.} 2006, \apjl,  636, L41

\bibitem[{Murray} {et~al.}(1997){Murray} {et~al.}]{m+97}
{Murray}, S.~S. {et~al.} 1997, \procspie, 3114, 11

\bibitem[{Ng} \& {Romani}(2007){Ng} \& {Romani}]{nr07}
{Ng}, C.-Y. \& {Romani}, R.~W. 2007, \apj, 660, 1357

\bibitem[Paczy\'{n}ski(1992)Paczy\'{n}ski]{p92}
Paczy\'{n}ski, B. 1992, Acta Astronomica, 42, 145

\bibitem[{Papoulis}(1991){Papoulis}]{papoulis91}
{Papoulis}, A. 1991, {Probability, random variables and stochastic processes},  3rd edn. (New York: McGraw-Hill)

\bibitem[{Patel} {et~al.}(2001){Patel}, {Kouveliotou}, {Woods}, {Tennant},  {Weisskopf}, {Finger}, {G{\" o}{\u g}{\" u}{\c s}}, {van der Klis}, \&  {Belloni}]{pkw+01}
{Patel}, S.~K., {et al.} 2001, \apjl, 563, L45

\bibitem[{Rothschild} {et~al.}(1994){Rothschild}, {Kulkarni}, \&  {Lingenfelter}]{rkl94}
{Rothschild}, R.~E., {Kulkarni}, S.~R., \& {Lingenfelter}, R.~E. 1994, \nat,  368, 432

\bibitem[{Skrutskie} {et~al.}(2006){Skrutskie}, {Cutri}, {Stiening},  {Weinberg}, {Schneider}, {Carpenter}, {Beichman}, {Capps}, {Chester},  {Elias}, {Huchra}, {Liebert}, {Lonsdale}, {Monet}, {Price}, {Seitzer},  {Jarrett}, {Kirkpatrick}, {Gizis}, {Howard}, {Evans}, {Fowler}, {Fullmer},  {Hurt}, {Light}, {Kopan}, {Marsh}, {McCallon}, {Tam}, {Van Dyk}, \&  {Wheelock}]{2mass}
{Skrutskie}, M.~F., {et al.} 2006, \aj, 131, 1163

\bibitem[{Tatematsu} {et~al.}(1990){Tatematsu}, {Fukui}, {Iwata}, {Seward}, \&  {Nakano}]{tfi+90}
{Tatematsu}, K., {Fukui}, Y., {Iwata}, T., {Seward}, F.~D., \& {Nakano}, M.  1990, \apj, 351, 157

\bibitem[{Taylor} {et~al.}(2003){Taylor}, {Gibson}, {Peracaula}, {Martin},  {Landecker}, {Brunt}, {Dewdney}, {Dougherty}, {Gray}, {Higgs}, {Kerton},  {Knee}, {Kothes}, {Purton}, {Uyaniker}, {Wallace}, {Willis}, \&  {Durand}]{tgp+03}
{Taylor}, A.~R., {et al.} 2003, \aj, 125, 3145

\bibitem[{Thompson} \& {Duncan}(1993){Thompson} \& {Duncan}]{td93}
{Thompson}, C. \& {Duncan}, R.~C. 1993, \apj, 408, 194

\bibitem[{Thompson} \& {Duncan}(1995){Thompson} \& {Duncan}]{td95}
---. 1995, \mnras, 275, 255

\bibitem[{Thompson} \& {Duncan}(1996){Thompson} \& {Duncan}]{td96}
---. 1996, \apj, 473, 322

\bibitem[{Vasisht} {et~al.}(1994){Vasisht}, {Kulkarni}, {Frail}, \&  {Greiner}]{vkfg94}
{Vasisht}, G., {Kulkarni}, S.~R., {Frail}, D.~A., \& {Greiner}, J. 1994, \apjl,  431, L35

\bibitem[{Vrba} {et~al.}(2000){Vrba}, {Henden}, {Luginbuhl}, {Guetter},  {Hartmann}, \& {Klose}]{vhl+00}
{Vrba}, F.~J., {Henden}, A.~A., {Luginbuhl}, C.~B., {Guetter}, H.~H.,  {Hartmann}, D.~H., \& {Klose}, S. 2000, \apjl, 533, L17

\bibitem[{Wachter} {et~al.}(2008){Wachter}, {Ramirez-Ruiz}, {Dwarkadas},  {Kouveliotou}, {Granot}, {Patel}, \& {Figer}]{wrrd+08}
{Wachter}, S., {Ramirez-Ruiz}, E., {Dwarkadas}, V.~V., {Kouveliotou}, C.,  {Granot}, J., {Patel}, S.~K., \& {Figer}, D. 2008, \nat, 453, 626

\bibitem[{Winkler} \& {Petre}(2007){Winkler} \& {Petre}]{wp07}
{Winkler}, P.~F. \& {Petre}, R. 2007, \apj, 670, 635

\bibitem[{Woods} \& {Thompson}(2006){Woods} \& {Thompson}]{wt06}
{Woods}, P.~M. \& {Thompson}, C. 2006, in Compact stellar X-ray sources, ed.  W.~{Lewin} \& M.~{van der Klis} (Cambridge, UK: Cambridge University Press),  547

\bibitem[{Zhang} \& {Harding}(2000){Zhang} \& {Harding}]{zh00}
{Zhang}, B. \& {Harding}, A.~K. 2000, \apjl, 535, L51

\end{thebibliography}


\end{document}